\documentclass[a4paper,onecolumn,11pt,]{quantumarticle}
\pdfoutput=1

\usepackage[numbers,sort&compress]{natbib}

\usepackage{graphicx}
\usepackage[dvipsnames,table,xcdraw]{xcolor}
\usepackage{amsmath,amssymb,dsfont}
\usepackage{braket}
\usepackage{bm}
\usepackage{enumerate}
\usepackage{amsthm}
\usepackage{comment}

\usepackage{enumerate}
\usepackage{physics2}
\usephysicsmodule{ab}
\usepackage{mathtools}

\usepackage{bbm}
\usepackage{enumitem}
\usepackage{multirow}
\usepackage{booktabs}
\usepackage{url}
\usepackage{xparse}
\usepackage[colorlinks,linkcolor=blue,citecolor=blue]{hyperref}
\usepackage{algorithm}
\usepackage{float}
\usepackage{algpseudocode}
\usepackage{listings}
\usepackage{xcolor}

\floatname{algorithm}{Pseudocode}

\lstdefinestyle{codeoutput}{
    language={},
    basicstyle=\ttfamily\small,
    backgroundcolor=\color{gray!10},
    frame=single,
    framerule=0.5pt,
    rulecolor=\color{gray!45},
    numbers=none,
    aboveskip=0.35\baselineskip,
    belowskip=0.5\baselineskip
}

\newcommand{\black}[1]{\textcolor{black}{#1}}

\newcommand{\D}{\mathbb{D}}

\newcommand{\mathI}{\mathcal{I}}
\newcommand{\mathJ}{\mathcal{J}}

\newcommand{\norm}[1]{\left\lVert #1 \right\rVert}
\newcommand{\abs}[1]{\left\lvert #1 \right\rvert}

\newtheorem{prop}{Proposition}
\newtheorem{lemma}{Lemma}

\usepackage{tikz}
\usepackage{quantikz}
\tikzset{octrlsquare/.style={shape=rectangle,draw,inner sep=0.8pt,minimum width=1.1ex,minimum height=1.1ex}}

\begin{document}
\title{pygridsynth: a fast numerical tool for ancilla-free Clifford+$T$ synthesis}

\author{Shuntaro Yamamoto}\email{shun0923@g.ecc.u-tokyo.ac.jp}
\affiliation{Department of Mathematical Informatics, Graduate School of Information Science and Technology, The University of Tokyo, Tokyo 113-8656, Japan}

\author{Nobuyuki Yoshioka}
\email{ny.nobuyoshioka@gmail.com}
\affiliation{\mbox{International Center for Elementary Particle Physics, The University of Tokyo, Tokyo 113-0033, Japan}}

\begin{abstract}
    We present \texttt{pygridsynth}, an open-source Python library for ancilla-free approximate Clifford+$T$ synthesis that runs in $O(\log(1/\epsilon))$ for precision $\epsilon$.
    For $n=1, 2$ qubits, the library builds upon established efficient and high-precision synthesis routines, such as nearly optimal $Z$-rotation synthesis and magnitude approximation. 
    For $n\ge 3$ qubits, we introduce a partial-decomposition technique that generalizes the magnitude approximation, reducing constant factors in the $T$-count as $(\frac{21}{8}\cdot 4^n - \frac{9}{2}\cdot 2^n + 9)\log_2(1/\epsilon) + o(\log(1/\epsilon))$.
    The package also exposes a mixed-synthesis workflow that approximates target unitary channels by probabilistic mixtures of Clifford+$T$ circuits, for which we empirically find that the synthesis error is reduced from $\epsilon$ to $\epsilon^2/(2n)$.
    Taken together, these features make \texttt{pygridsynth} a Python-native platform for high-precision Clifford$+T$ synthesis and for benchmarking unitary and mixed synthesis strategies on multi-qubit instances.
\end{abstract}

\maketitle

\section{Introduction}\label{sec:intro}

Fault-tolerant quantum computation requires compiling logical operations into a discrete universal gate set.
Among the most widely studied choices is the Clifford+$T$ gate set, in which Clifford operations are comparatively inexpensive while non-Clifford $T$ gates are typically mediated through magic-state cultivation and/or distillation~\cite{BravyiKitaev2004,BravyiHaah2012, OGormanCampbell2017}.
One of the most well-known methods is the Solovay-Kitaev algorithm~\cite{Kitaev2002, Dawson2006}, which states that finite number of $n$-qubit gates approximates arbitrary unitaries to accuracy $\epsilon$ using a number of gates polylogarithmic in $1/\epsilon$, if it generates a dense subgroup of SU($2^n$). Although its broad applicability is theoretically appealing, its practical performance is known to be far from optimal; in order to minimize the overhead of executing the quantum circuit, efficient synthesis over the Clifford+$T$ gate set is a problem of central importance.

A fundamental task in this context is \emph{approximate unitary synthesis}: given a target unitary operator and an error tolerance $\epsilon > 0$, find a circuit over the Clifford+$T$ gate set that approximates the target using as few $T$ gates as possible.
For the practically important case of Pauli rotations, which is equivalent to single-qubit Z-axis rotations $R_Z(\phi) = e^{-i\phi Z/2}$ up to Clifford conjugation that can be constructed efficiently, the problem has been thoroughly studied. Exact synthesis over rings such as $\mathbb{Z}[1/\sqrt{2},i]$ is well understood~\cite{Kliuchnikov2013fast,Matsumoto2008Representation,Giles2013Remarks},
and approximate unitary synthesis has progressed for near-optimal ancilla-free Pauli rotation gates~\cite{Ross2016optimal} which finds $3 \log_2(1/\epsilon) + O(\log\log(1/\epsilon))$ $T$ gates in computation time of $O(\log(1/\epsilon))$.
While information theoretic argument suggest that the lower bound on the T count for synthesizing general SU(2) gate is also $3 \log_2(1/\epsilon)$, there is no known efficient algorithm at the moment; the best algorithm with polylogarithmic computation time consumes $T$ gates of $7 \log_2(1/\epsilon)$ per gate~\cite{Kliuchnikov2023shorterquantum}, while the algorithm that achieves the lower bound requires polynomial runtime~\cite{morisaki2025optimal}.

The situation is less mature for Clifford+$T$ synthesis of general multi-qubit unitaries. Although there is a rich literature for decomposing generic multi-qubit unitaries into one- and two-qubit gates~\cite{Shende2006synthesis, Bergholm2005uniformly, Shende2004minimal,Vatan2004optimal, Krol2024beyond}, it is not even clear whether one can achieve the lower  bound on $T$-count of $\frac{4^n-1}{\log_2(4^n-2)}\log_2(1/\epsilon)$ which follows from the standard volume argument for ancilla-free unitary synthesis~\cite{harrow2002efficient}. Advancements are made for multi-controlled single-qubit gates~\cite{gosset2025multi, yamazaki2026multi}, while unitary synthesis of general multi-qubit gates is far less explored, not to mention the scarcity of synthesis tools.

A complementary recent direction is \emph{mixed} synthesis, in which a target unitary channel is approximated not by a single Clifford+$T$ circuit but by a convex mixture of several synthesized circuits.
It was originally pointed out independently by Hastings~\cite{Hastings2017turning} and Campbell~\cite{Campbell2017shorter} that such a mixture yields quadratic suppression of coherent errors for Pauli rotations, and even cubic to hexic suppression when {\it adaptive} synthesis utilizing ancilla with feedback operation is allowed~\cite{Kliuchnikov2023shorterquantum, Yoshioka2025error}. It is recognized that such an advantage could be present also in multi-qubit unitaries, with the lower bound of the accuracy for mixed synthesis being $\epsilon^2/2^n$ for qubit count $n$~\cite{akibue2024probabilistic}. 
Yet, this requires mixed synthesis over exponentially many unitaries with precisely tuned errors, and thus
it is not clear whether such a factor can be achieved under Clifford+$T$ synthesis.

As fault-tolerant quantum computing moves closer to practical realization, the need for high-performance \emph{transpilers}---compilers that translate high-level quantum circuits into hardware-native, fault-tolerant gate sequences---becomes increasingly pressing.
Gate synthesis lies at the core of such transpilers, and must be both fast and accessible to be useful at scale.
A Python-native implementation is particularly valuable in this regard, as it enables seamless integration with the existing ecosystem of quantum programming frameworks and facilitates rapid prototyping, benchmarking, and deployment within standard scientific workflows.

In this paper, we introduce \texttt{pygridsynth}, an open-source
publicly available on GitHub~\cite{pygridsynth} and PyPI~\cite{pygridsynth},
for approximate gate synthesis over the Clifford$+T$ gate set.
Our main contributions are as follows:
\begin{itemize}
    \item \textbf{High-precision implementation:} We implement all synthesis algorithms using the \texttt{mpmath} library for arbitrary-precision arithmetic.
          This enables decomposition of unitaries into Clifford+$T$ gates with extremely high precision, handling error tolerances as small as $\epsilon = 10^{-100}$ while maintaining numerical stability.

    \item \textbf{Multi-qubit synthesis with Block ZXZ decomposition:} We extend synthesis to three-qubit and larger systems using recursive block decomposition.
          For $n$-qubit systems with $n \geq 3$, we employ block ZXZ decomposition to break down unitaries into components that can be efficiently approximated using magnitude approximation~\cite{Kliuchnikov2023shorterquantum}.
          This strategy leverages magnitude approximation (achieving $T$-count $O(\log(1/\epsilon))$ for $X$-axis rotations) extensively, resulting in better $T$-count scaling than using only GridSynth approximations.

    \item \textbf{Partial decomposition for $T$-count reduction:} We introduce a partial decomposition strategy that leaves diagonal phase matrices undecomposed during multi-qubit synthesis.
          These diagonal matrices are absorbed into subsequent operations in the recursive structure, reducing the constant factor in the $T$-count complexity by avoiding unnecessary decomposition of intermediate phase matrices.

    \item \textbf{Multi-qubit mixed synthesis validation:} We demonstrate that mixed synthesis~\cite{Campbell2017shorter,Hastings2017turning,akibue2024probabilistic,Kliuchnikov2023shorterquantum,Yoshioka2025error} can be successfully applied to multi-qubit systems.
          Starting from Clifford+$T$ decompositions obtained through our multi-qubit synthesis algorithm, we numerically demonstrate that the theoretical prediction of quadratic error suppression holds in practice for multi-qubit systems, with a bonus factor (from $\epsilon$ to $\epsilon^2/(2n)$).
\end{itemize}

\black{Table~\ref{tab:$T$-count} summarizes the $T$-count for each synthesis method, where $\epsilon$ denotes the diamond-norm error tolerance. Throughout this paper, unless otherwise indicated, we omit subleading terms of  $o(\log(1/\epsilon))$ and report only the leading contribution to the asymptotic $T$-count.}
\begin{table}[htbp]
    \centering
    \caption{$T$-count for each synthesis method, where $\epsilon$ denotes the diamond-norm error tolerance. \black{The subleading term of $o(\log(1/\epsilon))$ is excluded.}}
    \label{tab:$T$-count}
    \begin{tabular}{ll}
        \hline\hline
        Method                                                                                     & $T$-count                                                                                            \\
        \hline
        Single-qubit synthesis (Sec.~\ref{sec:single-qubit},~\cite{Kliuchnikov2023shorterquantum}) & $\black{7\log_2(1/\epsilon)}$                                                                          \\
        Two-qubit synthesis (Sec.~\ref{sec:multi-qubit},~\cite{Kliuchnikov2023shorterquantum})     & $\black{33\log_2(1/\epsilon)}$                                                                         \\
        Multi-qubit synthesis, $n \geq 3$ qubits (Sec.~\ref{sec:multi-qubit})                      & $\black{\left(\frac{21}{8} \cdot 4^n - \frac{9}{2} \cdot 2^n + 9\right)\log_2(1/\epsilon)}$            \\
        \hline
        Mixed synthesis, single-qubit (Sec.~\ref{sec:mixed})                                       & $\black{\frac{7}{2}\log_2(1/\epsilon)}$                                                                \\
        Mixed synthesis, two-qubit (Sec.~\ref{sec:mixed})                                          & $\black{\frac{33}{2}\log_2(1/\epsilon)}$                                                               \\
        Mixed synthesis, $n \geq 3$ qubits (Sec.~\ref{sec:mixed})                                  & $\black{\left(\frac{21}{16} \cdot 4^n - \frac{9}{4} \cdot 2^n + \frac{9}{2}\right)\log_2(1/\epsilon)}$ \\
        \hline\hline
    \end{tabular}
\end{table}

The remainder of this paper is organized as follows.
Section~\ref{sec:gridsynth} describes the GridSynth algorithm for synthesizing $Z$-axis rotations, which is a fundamental building block for all synthesis methods.
Section~\ref{sec:single-qubit} explains single-qubit unitary synthesis, which combines Euler decomposition, magnitude approximation, and GridSynth.
Section~\ref{sec:multi-qubit} presents our proposed multi-qubit synthesis framework, including two-qubit decomposition, partial decomposition, and recursive decomposition for three-qubit and larger systems.
Section~\ref{sec:mixed} describes mixed synthesis, which approximates unitaries using probabilistic mixtures of Clifford+$T$ circuits.
Section~\ref{sec:experiments} presents numerical experiments validating the theoretical predictions.
Finally, Section~\ref{sec:conclusion} summarizes our contributions and discusses future directions.

\section{Single-qubit Pauli rotation gate synthesis}\label{sec:gridsynth}

\black{This section reviews synthesis of single-qubit $Z$-axis rotations $e^{-i\phi Z/2}$ in the Clifford+$T$ gate set, which serves as a fundamental building block later in the paper.
    Section~\ref{sec:gridsynth-background} collects the mathematical background: the Clifford+$T$ gate set and the synthesis goal for $Z$-rotations, the diamond-norm error metric, and the ring $\D[\omega]$.
    The algorithm that constructs an \emph{approximate} unitary meeting a prescribed error tolerance is known as \emph{GridSynth}~\cite{Ross2016optimal,Selinger2015efficient}; we describe it concretely in Section~\ref{sec:gridsynth-algorithm}.
    Finally, Section~\ref{sec:gridsynth-domega} treats \emph{exact synthesis}: rewriting the GridSynth output, which is stored as a $2\times 2$ unitary over $\D[\omega]$, into an explicit finite product of Clifford and $T$ gates, without introducing any further approximation error.}
The implementation used in this work is based on the reference implementation by Selinger and Ross~\cite{Selinger2018newsynth}.

\black{We find it beneficial for the readers to indicate an example usage of the library. The function \texttt{gridsynth} searches over $\D[\omega]$ and returns a symbolic unitary; \texttt{gridsynth\allowbreak\_gates} runs the same search and then builds a Clifford+$T$ circuit.
    The optional flag \texttt{up\_to\_phase} controls whether global phase is left unspecified: if \texttt{True}, a global phase may remain (synthesis is only up to global phase); if \texttt{False}, leftover phase is expanded into explicit gates (e.g.\ $W$) so the circuit fixes the overall phase.
    Source code~\ref{lst:gridsynth-api} shows a typical \texttt{gridsynth\allowbreak\_gates} call.}

\begin{lstlisting}[caption={GridSynth: \texttt{gridsynth\_gates}},label={lst:gridsynth-api}]
import mpmath
from pygridsynth.gridsynth import gridsynth_gates

theta, eps = mpmath.pi / 3, mpmath.mpf("1e-6")
gate_str = gridsynth_gates(theta=theta, epsilon=eps)
print(gate_str)
\end{lstlisting}
\noindent\textbf{Output (illustrative).}
\begin{lstlisting}[style=codeoutput]
HTSHTHTHTSHTHTHTHTSHTSHTHTSHTSHTHTHTSHTSHTSHTSHTHTHTHTSHT...
\end{lstlisting}

\subsection{Mathematical Background}\label{sec:gridsynth-background}

\subsubsection{Clifford+$T$ gate set}

\black{Recall that the single-qubit Clifford group is generated by the Hadamard gate~$H$, the phase gate~$S$, and the scalar phase $\omega = e^{i\pi/4}$. By adjoining the non-Clifford $T$~gate, one obtains a universal gate set for quantum computing with a single qubit:}
\begin{equation}
    \omega = e^{i\pi/4},\qquad
    H = \frac{1}{\sqrt{2}}\begin{pmatrix} 1 & 1 \\ 1 & -1 \end{pmatrix},\qquad
    S = \begin{pmatrix} 1 & 0 \\ 0 & i \end{pmatrix},\qquad
    T = \begin{pmatrix} 1 & 0 \\ 0 & e^{i\pi/4} \end{pmatrix}.
\end{equation}
\black{The basic synthesis problem treated here is to approximate an arbitrary $Z$-axis rotation}
\begin{equation}
    R_Z(\theta) = e^{-i\theta Z/2} = \begin{pmatrix} e^{-i\theta/2} & 0 \\ 0 & e^{i\theta/2} \end{pmatrix}
\end{equation}
\black{by a unitary realized as a Clifford+$T$ circuit, up to a prescribed tolerance $\epsilon > 0$ in the diamond norm (defined in Section~\ref{sec:gridsynth-diamond}).
    A foundational characterization~\cite{Kliuchnikov2013fast} is that a single-qubit unitary can be implemented \emph{exactly} over Clifford+$T$ (up to global phase) if and only if all of its matrix entries lie in the ring $\D[\omega] = \mathbb{Z}[1/\sqrt{2},i]$; we recall the algebraic structure of $\D[\omega]$ and its use in search and exact synthesis below.}

\subsubsection{Error metric: diamond norm}\label{sec:gridsynth-diamond}


Throughout this work, we measure the error tolerance between unitaries using the \emph{diamond norm} (also known as the completely bounded trace norm), which provides a worst-case error measure for quantum operations.
The diamond norm between two unitaries $U$ and $V$ is defined as:
\begin{equation}
    \norm{U - V}_\diamond = \max_{\rho} \norm{(U \otimes I)\rho(U^\dagger \otimes I) - (V \otimes I)\rho(V^\dagger \otimes I)}_1,
\end{equation}
where the maximum is taken over all density operators $\rho$ on the system plus an ancilla, and $\norm{\cdot}_1$ denotes the trace norm.
This metric is particularly important for quantum circuit synthesis because it captures the worst-case error over all possible input states, including entangled states with ancillas.

The diamond norm satisfies several important properties:
\begin{itemize}
    \item It provides an upper bound on the difference in measurement probabilities: if $\norm{U - V}_\diamond \leq \epsilon$, then for any input state $\rho$ and any measurement, the difference in outcome probabilities is at most $\epsilon$.
    \item It is stable under composition: if $\norm{U_1 - V_1}_\diamond \leq \epsilon_1$ and $\norm{U_2 - V_2}_\diamond \leq \epsilon_2$, then $\norm{U_2 U_1 - V_2 V_1}_\diamond \leq \epsilon_1 + \epsilon_2$.
    \item For unitary operators, the diamond norm can be computed efficiently via semidefinite programming with the Choi representation or through eigenvalue analysis.
\end{itemize}

All error tolerances $\epsilon$ in this work are specified in terms of the diamond norm.
That is, when we say a unitary $U$ is approximated by $\tilde{U}$ with error tolerance $\epsilon$, we mean $\norm{U - \tilde{U}}_\diamond \leq \epsilon$.

\subsubsection{Complex ring}
The algorithm operates in the ring $\D[\omega]$, where $\omega = e^{i\pi/4}$ is the eighth root of unity.
\black{Algebraically, every $u \in \D[\omega]$ can be written as $u = a\omega^3 + b\omega^2 + c\omega + d$ with $a,b,c,d \in \D$, where $\D = \mathbb{Z}\bigl[\tfrac{1}{2}\bigr] = \bigl\{ m/2^{\ell} : m \in \mathbb{Z},\; \ell \in \mathbb{Z}_{\geq 0} \bigr\}$.
    We write $\mathbb{Z}[\omega]=\{a\omega^3+b\omega^2+c\omega+d : a,b,c,d\in\mathbb{Z}\}$, $\mathbb{Z}[\sqrt{2}]=\{a+b\sqrt{2}: a,b\in\mathbb{Z}\}$, and $\D[\sqrt{2}]=\{a+b\sqrt{2}: a,b\in\D\}$.
    For Diophantine search and exact synthesis, implementations store elements $u \in \D[\omega]$ in the equivalent form $u = z/\sqrt{2}^k$ with $z \in \mathbb{Z}[\omega]$ and $k \geq 0$ minimal (the least denominator exponent).}

\black{
    Complex conjugation of the embedded complex number is denoted $u^\dagger$.
    For $u = a\omega^3 + b\omega^2 + c\omega + d$ with $a,b,c,d \in \D$, it sends $(a,b,c,d) \mapsto (-c,-b,-a,d)$ in the $(\omega^3,\omega^2,\omega,1)$ basis.
    The \emph{$\sqrt{2}$-conjugation}, which we denote $({}^\bullet)$, acts on $\mathbb{Z}[\omega]$ by the coordinate rule below; the same formula applies when $a,b,c,d \in \D$:
}
\begin{equation}
    (a\omega^3 + b\omega^2 + c\omega + d)^\bullet = -a\omega^3 + b\omega^2 - c\omega + d.
\end{equation}
\black{On $\mathbb{Z}[\sqrt{2}]$ and $\D[\sqrt{2}]$, $({}^\bullet)$ is given by}
\begin{equation}
    (a + b\sqrt{2})^\bullet = a - b\sqrt{2}.
\end{equation}

For the approximation of $Z$-axis rotations $e^{-i\phi Z/2}$, the epsilon region $R_\epsilon(\phi)$ is defined as:
\begin{equation}
    R_\epsilon(\phi) = \{u \in \mathbb{C} : |u| \leq 1 \text{ and } \mathrm{Re}(u^\dagger e^{-i\phi/2}) \geq 1 - \sqrt{1-\epsilon^2}/2\}.
\end{equation}
This region constrains the approximation accuracy in the complex plane, ensuring that the diamond norm error between the target rotation and its approximation is bounded by $\epsilon$.

\subsection{Algorithm Description}\label{sec:gridsynth-algorithm}

Given a rotation angle $\phi$ and error tolerance $\epsilon$, the GridSynth algorithm proceeds as follows~\cite{Ross2016optimal,Selinger2015efficient}:

\begin{enumerate}
    \item \textbf{Set up the constraint regions:}
          \begin{itemize}
              \item Let $A = R_\epsilon(\phi)$ be the $\epsilon$-region defined above, which constrains the approximation accuracy.
              \item Let $B$ be the unit disk in the complex plane.
          \end{itemize}

    \item \textbf{Enumerate solutions to the scaled grid problem:}
          Solve the two-dimensional grid problem (TDGP)~\cite{Ross2016optimal,Selinger2015efficient}: enumerate $u \in \mathbb{D}[\omega]$ such that $u \in A$ and $u^\bullet \in B$, in the order used by the reference implementation (increasing least denominator exponent in the $\D[\omega]$ normal form $u=z/\sqrt{2}^k$).

    \item \textbf{\black{Validate the candidates:}} For each \black{candidate} $u$ \black{from the TDGP in step~2,}
          \begin{enumerate}
              \item Let $\xi = 1 - u^\dagger u \in \D[\sqrt{2}]$, and write $\xi^\bullet \xi = \frac{n}{2^\ell}$, where $n \in \mathbb{Z}$ and $\ell \geq 0$ is minimal.

              \item Attempt to find a prime factorization of $n$.
                    If $n \neq 0$ but no prime factorization is found, skip step (c) and continue with the next \black{candidate}~$u$.

              \item Solve the diophantine equation $t^\dagger t = \xi$ to find $t$.
                    If a solution $t$ exists, go to step 4; otherwise, continue with the next \black{candidate}~$u$.
          \end{enumerate}

    \item \textbf{Construct the unitary:}
          Define $U = \begin{pmatrix} u & -t^\dagger \\ t & u^\dagger \end{pmatrix}$ (with $u, t$ as found above), let $U' = T U T^\dagger$, and keep whichever of $U,U'$ has the smaller $T$-count as the $\D[\omega]$-encoded candidate passed to exact synthesis.
          Output this candidate and stop.
\end{enumerate}

\black{\medskip It is beneficial to describe the
    \noindent\textit{Two-dimensional grid problem (TDGP) solved in the second step.}
    Recall $\mathbb{Z}[\omega]\subseteq\mathbb{C}$. Following Ross and Selinger~\cite{Ross2016optimal,Selinger2015efficient}, it is convenient to identify $\mathbb{C}\cong\mathbb{R}^2$ and to regard $\mathbb{Z}[\omega]$ as a subset of the plane.
    For $B\subseteq\mathbb{R}^2$, the \emph{complex grid for $B$} is}
\begin{equation}
    \mathrm{Grid}(B) = \{\, u \in \mathbb{Z}[\omega] \mid u^\bullet \in B \,\}.
\end{equation}
\black{If $B$ is bounded, convex, and has non-empty interior, then $\mathrm{Grid}(B)$ is discrete and infinite~\cite{Ross2016optimal}.
    Given $A,B\subseteq\mathbb{R}^2$, the TDGP is to find $u\in\mathbb{Z}[\omega]$ with $u\in A$ and $u^\bullet\in B$, i.e.\ $u\in A\cap\mathrm{Grid}(B)$: one seeks points in the intersection of $A$ with the grid determined by~$B$.
    Ross and Selinger give an efficient procedure that enumerates all such solutions for bounded convex $A,B$ with non-empty interior~\cite{Ross2016optimal,Selinger2015efficient}; GridSynth instantiates $A=R_\epsilon(\phi)$ and $B$ the unit disk, and step~2 above runs this search with the scaling and ordering from~\cite{Ross2016optimal,Selinger2015efficient}.}

The algorithm systematically searches for solutions with increasing $k$, ensuring that the first valid solution found has minimal or near-minimal $T$-count.
The expected runtime is $O(\mathrm{polylog}(1/\epsilon))$ under mild number-theoretic assumptions.

\subsubsection{$T$-count Analysis}

The GridSynth algorithm achieves near-optimal $T$-count for $Z$-axis rotations.
For a rotation angle $\phi$ and error tolerance $\epsilon$, the resulting Clifford+$T$ decomposition has a $T$-count of $\black{3\log_2(1/\epsilon) + o(\log(1/\epsilon))}$.
This means that the number of $T$ gates required scales logarithmically with the inverse of the error tolerance, with a constant factor of approximately 3.
The algorithm finds solutions with $T$-count that is within an additive constant of the optimal value.

\subsection{\black{Exact synthesis: from $\D[\omega]$ to Clifford+$T$ gates}}\label{sec:gridsynth-domega}

\black{GridSynth outputs the approximating rotation as a $2\times 2$ special unitary whose matrix entries lie in $\D[\omega]$.
    Such matrices are already elements of the abstract Clifford+$T$ group; the remaining task is \emph{exact synthesis}---to expand them as a concrete finite product of Clifford and $T$ gates~\cite{Kliuchnikov2013fast,Matsumoto2008Representation,Giles2013Remarks}.
    This step changes only the representation (from ring arithmetic to a gate list) and does not alter the unitary beyond the usual global-phase conventions.}

The routine \texttt{decompose\_domega\_unitary} proceeds as follows:
\begin{enumerate}
    \item \black{\textbf{Denominator loop ($k>0$).} While the least denominator exponent $k$ is positive, multiply the current $\D[\omega]$-symbol on the left by a short Clifford+$T$ prefix (built from $H$, $T$, and $S$) so that $k$ decreases by one; the prefixes are chosen systematically from the ring-theoretic data of the matrix.}

    \item \black{\textbf{Tail at $k=0$.} Once $k=0$, a short one-qubit Clifford tail clears the remaining bookkeeping (including $T$, $S_X$, $S$, and phase content, up to an optional global phase).}

    \item \black{\textbf{Normal form.} Finally, the gate list is simplified and rewritten into a fixed conventional form so that equivalent circuits are presented in the same way and redundant gates are removed where possible.}
\end{enumerate}

\black{Each left-multiplication preserves the encoded unitary; the loop terminates with $k=0$, after which the tail reductions drive the symbolic matrix to the identity.
    Because every Clifford+$T$ group element represented over $\D[\omega]$ admits this expansion, the output circuit implements the same unitary as the GridSynth solution, with no additional approximation error beyond global-phase bookkeeping.}

\section{Single-Qubit Synthesis}\label{sec:single-qubit}

\black{This subsection addresses the \emph{approximate single-qubit unitary synthesis} problem: given an arbitrary target $U \in \mathrm{SU}(2)$ and a tolerance $\epsilon > 0$, find a Clifford+$T$ circuit whose unitary agrees with $U$ up to diamond-norm error at most~$\epsilon$.
    The pipeline of Kliuchnikov~\textit{et~al.}~\cite{Kliuchnikov2023shorterquantum}, as implemented in \texttt{pygridsynth}, solves this problem; we record prior art rather than new algorithmic ideas.
    The subsections that follow spell out Euler decomposition, magnitude approximation, and GridSynth; here we only summarize the end-to-end workflow.}

\begin{algorithm}[b]
    \caption{Roadmap for single-qubit Clifford+$T$ synthesis (Section~\ref{sec:single-qubit}).}
    \label{alg:single-qubit-roadmap}
    \begin{algorithmic}[1]
        \Require $U \in \mathrm{SU}(2)$, tolerance $\epsilon$
        \Ensure Clifford+$T$ circuit and implemented unitary within $\epsilon$ of $U$ (diamond norm)
        \State Euler-decompose $U$ into $Z$--$X$--$Z$ data $(\phi,\phi_1,\theta,\phi_2)$.
        \State Approximate $e^{-i\theta X/2}$ with magnitude approximation at tolerance $\epsilon/3$.
        \State Merge residual $Z$-rotations with $\phi_1,\phi_2$; approximate each combined $Z$-factor with GridSynth at tolerance $\epsilon/3$.
        \State Concatenate the Clifford+$T$ blocks; handle the global phase $e^{i\phi}$ if needed.
    \end{algorithmic}
\end{algorithm}

\black{The entry point \texttt{approximate\_one\_qubit\_unitary} implements Pseudocode~\ref{alg:single-qubit-roadmap}; source code~\ref{lst:single-qubit-api} shows a minimal call.
    It returns \texttt{circuit} (gate list) and \texttt{U\_approx} (implemented unitary, up to the library's global-phase conventions).}
\begin{lstlisting}[caption={Single-qubit synthesis: \texttt{approximate\_one\_qubit\_unitary}},label={lst:single-qubit-api}]
import mpmath as mp
import numpy as np
from pygridsynth.mymath import random_su
from pygridsynth.unitary_approximation import approximate_one_qubit_unitary

np.random.seed(0)
U = random_su(1)
circuit, U_approx = approximate_one_qubit_unitary(
    unitary=U,
    epsilon=mp.mpf("1e-5"),
    wires=[0],
)
print("len(circuit):", len(circuit))
print("U_approx:", U_approx)
\end{lstlisting}
\noindent\textbf{Output (illustrative).}
\begin{lstlisting}[style=codeoutput]
len(circuit): 333
U_approx: [(-0.583 - 0.346j)   (0.642 - 0.359j)]
          [(-0.642 - 0.359j)  (-0.583 + 0.346j)]
\end{lstlisting}

\subsection{Euler Decomposition}

Any single-qubit unitary matrix $U \in \mathrm{U}(2)$ can be decomposed using the Euler decomposition:
\begin{equation}
    U = e^{i\phi} e^{-i\phi_1 Z/2} e^{-i\theta X/2} e^{-i\phi_2 Z/2},
\end{equation}
where $\phi, \phi_1, \phi_2 \in [-\pi, \pi)$ and $\theta \in [0, \pi]$.

The decomposition parameters are computed from the unitary matrix elements as follows:
\begin{align}
    \phi   & = \frac{1}{2} \arg\det(U),                             \\
    \theta & = 2 \arctan2\left(\abs{U_{2,1}}, \abs{U_{2,2}}\right), \\
    \phi_1 & = \psi_1 + \psi_2 - \arg\det(U) + \frac{\pi}{2},       \\
    \phi_2 & = \psi_1 - \psi_2 - \frac{\pi}{2},
\end{align}
where $\psi_1 = \arg(U_{2,2})$, $\psi_2 = \arg(U_{2,1})$, and $\arg(\cdot)$ denotes the complex argument in the range $[-\pi, \pi)$.
Here, $U_{i,j}$ denotes the $(i,j)$-th element of the unitary matrix $U$ (using 1-based indexing for matrix elements).

\subsection{Magnitude Approximation}

The key component of single-qubit synthesis is the magnitude approximation algorithm, implemented in \texttt{magnitude\_approximate}.
This function approximates the rotation angle $\theta$ for the $X$-axis rotation $e^{-i\theta X/2}$ \black{up to residual Z-axis rotation as follows}~\cite{Kliuchnikov2023shorterquantum}:


\begin{prop}[Magnitude approximation condition]
    Let $\theta \in [0, \pi/2]$ and $\epsilon \leq 2$ be a positive real number.
    Suppose that we have found a special unitary $V$ given by a Clifford+$T$ decomposition:
    \begin{equation}
        V = g_1 \ldots g_n = \begin{pmatrix} u & -t^\dagger \\ t & u^\dagger \end{pmatrix}
    \end{equation}
    such that $|u|$ belongs to the interval $\{\cos(\theta'') : \theta'' \in [0, \pi/2], \abs{\theta'' - (-\theta/2)} \leq \delta\}$ for $\delta = \arcsin(\epsilon/2)$.
    Then unitary $V$ satisfies the inequality
    \begin{equation}
        \norm{V - e^{-i \phi'_1 Z / 2} e^{-i \theta X / 2} e^{-i \phi'_2 Z / 2}}_\diamond \leq \epsilon,
    \end{equation}
    for $\phi'_1$ and $\phi'_2$ defined by the equality $V = e^{-i\phi'_1 Z / 2} e^{-i\theta' X / 2} e^{-i\phi'_2 Z / 2}$ with $\theta' \in [0, \pi]$.
    We call such $V$ a magnitude $\epsilon$-approximation of $e^{-i\theta X / 2}$.
\end{prop}

The algorithm operates by finding elements in the ring $\D[\sqrt{2}] = \{(a + b\sqrt{2}) / \sqrt{2}^k : a, b, k \in \mathbb{Z}, k \geq 0\}$ whose magnitude approximates $\cos(\theta/2)$ within the specified error tolerance.

\subsubsection{$T$-count Analysis}

The magnitude approximation algorithm achieves efficient $T$-count for $X$-axis rotations.
For a rotation angle $\theta$ and error tolerance $\epsilon$, the resulting Clifford+$T$ decomposition has a $T$-count of $\log_2(1/\epsilon) + o(\log(1/\epsilon))$.
\black{Note that the constant factor is lower} than using GridSynth for $X$-axis rotations, which would require approximately three-fold larger number of $T$ gates.
The magnitude approximation method exploits the one-dimensional grid structure, resulting in a lower asymptotic complexity compared to two-dimensional grid problems.

\subsubsection{Algorithm Description}

Given a rotation angle $\theta$ and error tolerance $\epsilon$, the magnitude approximation algorithm proceeds as follows:

\begin{enumerate}
    \item \textbf{Set up the constraint interval:}
          \begin{itemize}
              \item Generate an epsilon interval $\mathI_\epsilon(\theta) = [\cos^2(-\theta/2 - \epsilon/2), \cos^2(-\theta/2 + \epsilon/2)]$ based on the target angle $\theta$ and error tolerance $\epsilon$.
              \item Let $\mathJ$ be the unit interval $[0, 1]$.
          \end{itemize}

    \item \textbf{Enumerate solutions to the scaled grid problem:}
          \begin{enumerate}
              \item \black{Solve the one-dimensional grid problem (ODGP)~\cite{Ross2016optimal,Selinger2015efficient}: enumerate $m \in \mathbb{Z}[\sqrt{2}]$ such that $m \in \mathI_\epsilon(\theta)$ and $m^\bullet \in \mathJ$, in the order used by the reference implementation (increasing least denominator exponent in the $\D[\sqrt{2}]$ normal form $m=r/\sqrt{2}^k$).}

              \item Solve the diophantine equation $u^\dagger u = m$ to find $u$.
                    If a solution $u$ exists, go to step 3; otherwise, continue with the next $m$.
          \end{enumerate}

    \item \textbf{\black{Validate the candidates:}} For each \black{candidate} $u$ \black{produced in step~2 (satisfying $u^\dagger u = m$ for the current grid point~$m$),}
          \begin{enumerate}
              \item Let $\xi = 1 - u^\dagger u \in \D[\sqrt{2}]$, and write $\xi^\bullet \xi = \frac{n}{2^\ell}$, where $n \in \mathbb{Z}$ and $\ell \geq 0$ is minimal.

              \item Attempt to find a prime factorization of $n$.
                    If $n \neq 0$ but no prime factorization is found, skip step (c) and continue with the next \black{candidate}~$u$.

              \item Solve the diophantine equation $t^\dagger t = \xi$ to find $t$.
                    If a solution $t$ exists, go to step 4; otherwise, continue with the next \black{candidate}~$u$.
          \end{enumerate}

    \item \textbf{Construct the approximation:}
          \black{Define $U = \begin{pmatrix} u & -t^\dagger \\ t & u^\dagger \end{pmatrix}$ (with $u, t$ as found above) as the $\D[\omega]$-encoded approximation.
          Output this candidate and stop.}
\end{enumerate}

\black{\medskip
    \noindent\textit{One-dimensional grid problem (ODGP).}
    Following Ross and Selinger~\cite{Ross2016optimal,Selinger2015efficient}, for $B\subseteq\mathbb{R}$ the \emph{real grid for $B$} is}
\begin{equation}
    \mathrm{grid}(B) = \{\, \alpha \in \mathbb{Z}[\sqrt{2}] \mid \alpha^\bullet \in B \,\}.
\end{equation}
\black{Elements of $\mathrm{grid}(B)$ are called \emph{grid points} when $B$ is clear from context~\cite{Ross2016optimal}.
Here we take $B=[y_0,y_1]$ to be a closed interval with $y_0<y_1$; then $\mathrm{grid}(B)$ is discrete and infinite~\cite{Ross2016optimal}.
For $A,B\subseteq\mathbb{R}$, the ODGP is to find $\alpha\in\mathbb{Z}[\sqrt{2}]$ with $\alpha\in A$ and $\alpha^\bullet\in B$, equivalently $\alpha\in A\cap\mathrm{grid}(B)$; one calls $\alpha\in A$ and $\alpha^\bullet\in B$ the \emph{grid constraints}~\cite{Ross2016optimal}.
Magnitude approximation uses $A=\mathI_\epsilon(\theta)$ and $B=\mathJ=[0,1]$ and runs the enumeration in step~2 with the scaling and ordering from~\cite{Ross2016optimal,Selinger2015efficient}.}

The algorithm systematically searches for solutions with increasing $k$, ensuring that the first valid solution found provides a magnitude $\epsilon$-approximation of \black{$e^{-i \theta X / 2}$}.

\subsection{Combining X and Z Rotations}

To synthesize an arbitrary single-qubit unitary, we combine magnitude approximation for $X$-axis rotations with GridSynth for $Z$-axis rotations.
The key insight is that by carefully managing the error allocation and combining residual $Z$-rotations, we can obtain an approximate unitary that can be exactly decomposed into Clifford+$T$ gates.

Given a target unitary $U$ and error tolerance $\epsilon$, the procedure is as follows:

\begin{enumerate}
    \item \textbf{Euler decomposition:}
          Decompose $U$ into $U = e^{i\phi} e^{-i\phi_1 Z/2} e^{-i\theta X/2} e^{-i\phi_2 Z/2}$, extracting parameters $\phi, \phi_1, \theta, \phi_2$.

    \item \textbf{Magnitude approximation of X-rotation:}
          Approximate $e^{-i\theta X/2}$ using magnitude approximation with error tolerance $\epsilon/3$, obtaining a magnitude $\epsilon/3$-approximation $V_{X}$.
          This gives us a unitary $V_X = e^{i\phi_1^r Z/2} e^{-i\theta' X/2} e^{i\phi_2^r Z/2}$ (up to a global phase), where $\theta'$ is close to $\theta$ and $\phi_1^r, \phi_2^r$ are residual $Z$-rotations introduced by the approximation.

    \item \textbf{Combining residual Z-rotations:}
          The residual $Z$-rotations $\phi_1^r$ and $\phi_2^r$ from the magnitude approximation are combined with the original $Z$-rotations from the Euler decomposition:
          \begin{align}
              \phi_1' & = \phi_1 - \phi_1^r, \\
              \phi_2' & = \phi_2 - \phi_2^r.
          \end{align}

    \item \textbf{GridSynth approximation of combined Z-rotations:}
          Approximate $e^{-i\phi_1' Z/2}$ and $e^{-i\phi_2' Z/2}$ using GridSynth with error tolerance $\epsilon/3$ each.
          \black{GridSynth supplies each $Z$-rotation in $\D[\omega]$ form, ready for exact synthesis into Clifford+$T$ gates.}

    \item \textbf{Constructing the final circuit:}
          Combine all components in the order $e^{-i\phi_1' Z/2} \cdot V_X \cdot e^{-i\phi_2' Z/2}$.
          \black{Each piece is first given as a $\D[\omega]$-representable unitary; exact synthesis turns every piece into a Clifford+$T$ circuit, so the full product is again a Clifford+$T$ circuit.
          Therefore, the final approximate unitary admits an exact Clifford+$T$ realization once these expansions are carried out.}
\end{enumerate}

By using error tolerance $\epsilon/3$ for each of the three approximation steps (left $Z$-rotation, $X$-rotation, and right $Z$-rotation), the total error accumulates to at most $\epsilon$ due to the subadditivity of the diamond norm under composition.

\subsection{$T$-count Analysis for Single-Qubit Synthesis}

The single-qubit synthesis process combines GridSynth (for $Z$-axis rotations) and magnitude approximation (for $X$-axis rotations).
Since Euler decomposition yields two $Z$-axis rotations and one $X$-axis rotation, and each rotation is approximated with error tolerance $\epsilon/3$, the total $T$-count is:
\begin{equation}
    \begin{aligned}
        T_{\text{1-qubit}} & = \black{2 \cdot 3\log_2\bigl(1/(\epsilon/3)\bigr) + \log_2\bigl(1/(\epsilon/3)\bigr) + o(\log(1/\epsilon))} \\
                           & = \black{7\log_2(1/\epsilon) + o(\log(1/\epsilon))}.
    \end{aligned}
\end{equation}
This means that the single-qubit synthesis requires $\black{7\log_2(1/\epsilon)}$ $T$ gates, which comes from two GridSynth approximations (each contributing $\black{3\log_2(1/\epsilon)}$ $T$ gates) and one magnitude approximation (contributing $\black{\log_2(1/\epsilon)}$ $T$ gates).

\section{Multi-Qubit Unitary Synthesis}\label{sec:multi-qubit}

\black{This section records how \texttt{pygridsynth} approximates arbitrary $n$-qubit unitaries over Clifford+$T$.
    Subsection~\ref{sec:two-qubit} is a review of the three-CNOT two-qubit template and its approximation pipeline (prior art).
    Subsection~\ref{sec:multi-qubit-recursive} presents our proposal for $n \geq 3$: recursive synthesis driven by the block ZXZ decomposition~\cite{Krol2024beyond}, together with demultiplexing and multiplexed Pauli rotation gates as detailed below.
    For multi-qubit unitary approximation, we employ a recursive decomposition strategy that breaks down an $n$-qubit unitary into smaller components that can be synthesized independently under allocated diamond-norm budgets.
    Source code~\ref{lst:multi-qubit-api} shows a minimal call to the $n$-qubit driver \texttt{approximate\_multi\_qubit\_unitary}.}

\begin{lstlisting}[caption={Multi-qubit synthesis: \texttt{approximate\_multi\_qubit\_unitary}},label={lst:multi-qubit-api}]
import mpmath as mp
from pygridsynth.multi_qubit_unitary_approximation import (
    approximate_multi_qubit_unitary,
)

U = mp.eye(8)
circuit, U_approx = approximate_multi_qubit_unitary(
    unitary=U,
    num_qubits=3,
    epsilon=mp.mpf("1e-5"),
)
print("gates:", len(circuit))
\end{lstlisting}
\noindent\textbf{Output (illustrative).}
\begin{lstlisting}[style=codeoutput]
gates: 1562
\end{lstlisting}

\subsection{Two-Qubit Unitary Approximation}\label{sec:two-qubit}

The two-qubit case serves as the base case for the recursive decomposition.
The function \texttt{approximate\_two\_qubit\_unitary} implements a decomposition that uses exactly \black{three} CNOT gates, which is optimal for two-qubit unitaries.

\subsubsection{Decomposition with Three CNOTs}\label{sec:decomposition-three-cnots}

The complete two-qubit decomposition uses three CNOT gates, which is optimal for arbitrary two-qubit unitaries~\cite{Shende2004minimal,Shende2004smaller}.
The decomposition is performed by \texttt{\_decompose\allowbreak\_two\allowbreak\_qubit\allowbreak\_unitary}, which implements the algorithm described below.
\black{The two-CNOT decomposition applies to those unitaries $U$ for which $\mathrm{tr}(\gamma(U))$ is real~\cite{Shende2004minimal}, i.e.\ the trace hypothesis in Proposition~V.2 (step~2 below fixes the explicit map~$\gamma$).}

The algorithm proceeds as follows, based on Proposition V.2~\cite{Shende2004minimal}:
\begin{enumerate}
    \item Transform the unitary as $U_2 = e^{i\pi/4} U \cdot \mathrm{CNOT}_{01}$.

    \item Compute $\gamma(U_2^T)$ to obtain a local equivalence invariant, where $\gamma(u) = u \sigma_y \otimes \sigma_y u^T \sigma_y \otimes \sigma_y$~\cite{Shende2004minimal}.

    \item Determine the phase parameter $\psi$ from $\gamma(U_2^T)$ to ensure the trace condition required by Proposition V.2~\cite{Shende2004minimal}.

    \item Construct $\Delta = \mathrm{CNOT}_{01} (I \otimes R_z(\psi)) \mathrm{CNOT}_{01}$.

    \item Apply the two-CNOT decomposition to $U_2 \Delta$ using \texttt{\_decompose\_with\_2\_cnots} to extract single-qubit unitaries $A, B, C, D$ and rotation angles $\theta, \phi$.
\end{enumerate}

The extraction of prefactors $A, B, C, D$ is performed by \texttt{\_extract\_SU2SU2\_prefactors}, which uses singular value decomposition and matrix reshaping to find the tensor product structure.

This decomposition yields a circuit structure of the form shown in Fig.~\ref{fig:two-qubit-decomposition} (top).
The circuit consists of single-qubit gates $A, B, C, D$ surrounding a core structure with three CNOT gates and the rotations $R_x(\theta)$, $R_z(\phi)$, and $R_z(-\psi)$.

\subsubsection{Approximation Process}\label{sec:approximation-process-two}

The approximation process for two-qubit unitaries follows the methodology proposed by Kliuchnikov~\textit{et~al.}~\cite{Kliuchnikov2023shorterquantum}:
\begin{enumerate}
    \item Decompose the unitary into the \black{three-CNOT} structure with single-qubit gates.
    \item Approximate the $R_x(\theta)$ and $R_z(\phi)$ rotations using \texttt{magnitude\_approximate} and decompose them.
    \item Approximate each single-qubit gate $A, B, C, D$ using \texttt{approximate\_one\_qubit\_unitary}.
    \item Combine all approximated gates into a single circuit.
\end{enumerate}

\begin{figure}[htbp]
    \centering
    \small
    \begin{quantikz}[column sep=0.4em, row sep=0.6em]
        \lstick{control} & \qw & \ctrl{1} & \gate{C} & \ctrl{1} & \gate{R_x} & \ctrl{1} & \gate{A} & \qw \\
        \lstick{target}   & \gate{R_z} & \targ{}  & \gate{D} & \targ{}  & \gate{R_z} & \targ{}  & \gate{B} & \qw
    \end{quantikz}
    \\[0.8em]
    $=$
    \begin{quantikz}[column sep=0.4em, row sep=0.6em]
        \lstick{control} & \qw & \ctrl{1} & \gate{C} & \ctrl{1} & \gate{R_z} & \gate{\mathrm{CT}} & \gate{R_z} & \ctrl{1} & \gate{A} & \qw \\
        \lstick{target}   & \gate{R_z} & \targ{}  & \gate{D} & \targ{}  & \gate{R_x} & \gate{\mathrm{CT}} & \gate{R_x} & \targ{}  & \gate{B} & \qw
    \end{quantikz}
    \\[0.8em]
    $=$
    \begin{quantikz}[column sep=0.4em, row sep=0.6em]
        \lstick{control} & \qw & \ctrl{1} & \gate{C'} & \ctrl{1} & \gate{\mathrm{CT}} & \ctrl{1} & \gate{A'} & \qw \\
        \lstick{target}   & \gate{R_z} & \targ{}  & \gate{D'} & \targ{} & \gate{\mathrm{CT}} & \targ{}  & \gate{B'} & \qw
    \end{quantikz}
    \caption{Two-qubit decomposition and absorption of residuals. (1)~Three-CNOT template (top): $R_z(-\psi)$ on the target; $(C,D)$ between CNOTs 1--2; $R_x(\theta)$ and $R_z(\phi)$ between CNOTs 2--3; $(A,B)$ after CNOT~3. (2)~After magnitude approximation, residual $R_z$ (control) and $R_x$ (target) appear around the Clifford+$T$ core. (3)~Residuals commute with CNOTs and merge into $C',D',A',B'$ and a simpler core.}
    \label{fig:two-qubit-decomposition}
\end{figure}
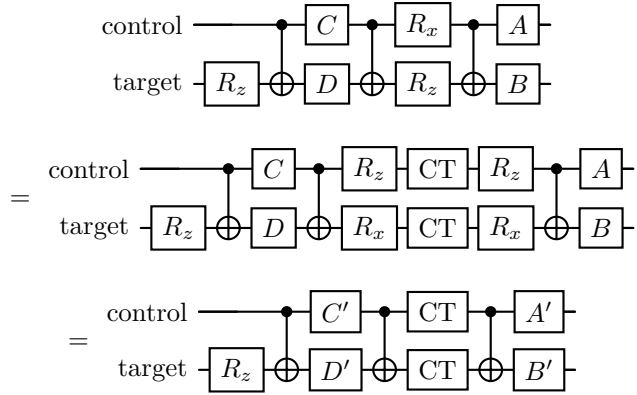

An important observation is that the residual $Z$ and $X$ rotations introduced by the magnitude approximation of $R_x(\theta)$ and $R_z(\phi)$ are single-qubit gates that commute with CNOT gates.
Specifically, $R_z$ rotations on the \black{control} qubit and $R_x$ rotations on the \black{target} qubit commute with CNOT gates.
This allows these residual rotations to be absorbed into the single-qubit gates $A, B, C, D$ through circuit simplification, as illustrated in Fig.~\ref{fig:two-qubit-decomposition} (middle $\to$ bottom).
This absorption reduces the overall gate count by eliminating redundant rotations.

\black{The function \texttt{approximate\allowbreak\_two\allowbreak\_qubit\allowbreak\_unitary} builds the three-CNOT template, approximates the core and tensor factors, and returns a Clifford+$T$ circuit.}

\subsubsection{$T$-count Analysis for Two-Qubit Synthesis}

The two-qubit synthesis process involves approximating multiple single-qubit gates and rotations.
For the complete decomposition with three CNOTs, we need to approximate:
\begin{itemize}
    \item Four single-qubit gates \black{$A', B', C', D'$}: Each requires $\black{7\log_2(1/\epsilon)}$ $T$ gates, contributing $4 \times \black{7\log_2(1/\epsilon)} = \black{28\log_2(1/\epsilon)}$ total.
    \item One $R_x(\theta)$ rotation: Requires $\black{\log_2(1/\epsilon)}$ $T$ gates via magnitude approximation.
    \item One $R_z(\phi)$ rotation: Requires $\black{\log_2(1/\epsilon)}$ $T$ gates via magnitude approximation.
    \item One additional $R_z(- \psi)$ rotation: Requires $\black{3\log_2(1/\epsilon)}$ $T$ gates via GridSynth.
\end{itemize}
Note that the residual $Z$ and $X$ rotations from magnitude approximation are absorbed into the single-qubit gates $A, B, C, D$, reducing the gate count.
The total $T$-count for two-qubit synthesis is:
\begin{equation}
    T_{\text{2-qubit}} = \black{28\log_2(1/\epsilon) + 2\log_2(1/\epsilon) + 3\log_2(1/\epsilon) + o(\log(1/\epsilon)) = 33\log_2(1/\epsilon) + o(\log(1/\epsilon))},
\end{equation}
which accounts for the four single-qubit approximations ($\black{28\log_2(1/\epsilon)}$), two magnitude approximations ($\black{2\log_2(1/\epsilon)}$), and one GridSynth approximation ($\black{3\log_2(1/\epsilon)}$).

\subsubsection{Partial Decomposition}\label{sec:partial-decomposition}

\black{This subsection explains synthesis that \emph{partially} decomposes a two-qubit unitary: intermediate steps leave some diagonal degrees of freedom unsynthesized rather than expanding them immediately to Clifford+$T$.}
\black{Such \emph{partial} factorizations are introduced because they reduce the $T$-count when the same two-qubit block appears inside multi-qubit ($n\ge 3$) synthesis; Sec.~\ref{sec:approx-decomposed-components} gives the detailed workflow.}
\black{As in source code~\ref{lst:two-qubit-partial}, \texttt{decompose\_partially=True} on \texttt{approximate\_two\_qubit\_unitary} keeps the unsynthesized phase block $\Delta_{\text{phase}}$ below explicit while the remaining gates are compiled to Clifford+$T$.}

\begin{lstlisting}[caption={Two-qubit synthesis with partial decomposition (\texttt{decompose\_partially=True})},label={lst:two-qubit-partial}]
from pygridsynth.unitary_approximation import approximate_two_qubit_unitary

circuit, U_approx = approximate_two_qubit_unitary(
    unitary=U,
    epsilon=eps,
    wires=[0, 1],
    decompose_partially=True,
)
\end{lstlisting}

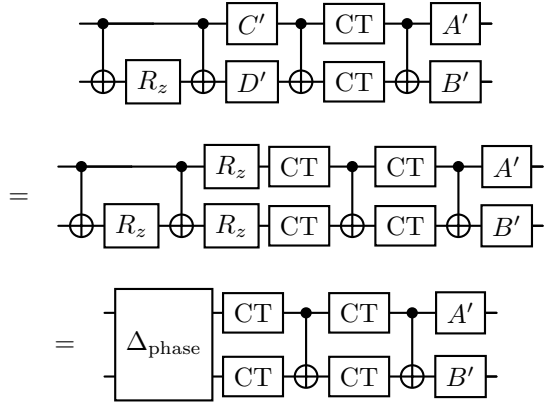
\begin{figure}[htbp]
    \centering
    \small
    \begin{quantikz}[column sep=0.4em, row sep=0.6em]
        \lstick{} & \ctrl{1} & \qw & \ctrl{1} & \gate{C'} & \ctrl{1} & \gate{\mathrm{CT}} & \ctrl{1} & \gate{A'} & \qw \\
        \lstick{} & \targ{} & \gate{R_z} & \targ{} & \gate{D'} & \targ{} & \gate{\mathrm{CT}} & \targ{} & \gate{B'} & \qw
    \end{quantikz}\\[0.8em]
    $=$
    \begin{quantikz}[column sep=0.4em, row sep=0.6em]
        \lstick{} & \ctrl{1} & \qw & \ctrl{1} & \gate{R_z} & \gate{\mathrm{CT}} & \ctrl{1} & \gate{\mathrm{CT}} & \ctrl{1} & \gate{A'} & \qw \\
        \lstick{} & \targ{} & \gate{R_z} & \targ{} & \gate{R_z} & \gate{\mathrm{CT}} & \targ{} & \gate{\mathrm{CT}} & \targ{} & \gate{B'} & \qw
    \end{quantikz}\\[0.8em]
    $=$
    \begin{quantikz}[column sep=0.4em, row sep=0.6em]
        \lstick{} & \gate[wires=2]{\Delta_{\text{phase}}} & \gate{\mathrm{CT}} & \ctrl{1} & \gate{\mathrm{CT}} & \ctrl{1} & \gate{A'} & \qw \\
        \lstick{} & & \gate{\mathrm{CT}} & \targ{} & \gate{\mathrm{CT}} & \targ{} & \gate{B'} & \qw
    \end{quantikz}
    \caption{\black{Partial decomposition of a two-qubit block. \emph{Top:} the absorbed form from Fig.~\ref{fig:two-qubit-decomposition} (bottom), with an additional leading $\mathrm{CNOT}_{01}$; primed blocks $C',D',A',B'$, three CNOTs in the core, and Clifford+$T$ cores ($\mathrm{CT}$). \emph{Middle:} $C'$ and $D'$ expanded via~\eqref{eq:euler-CD-prime} into $R_z$ and $\mathrm{CT}$ layers on each wire; the leading $\mathrm{CNOT}_{01}$ and the rest of the CNOT pattern are unchanged. \emph{Bottom:} $\Delta_{\text{phase}}$~\eqref{eq:delta-phase-bank} absorbs the leading $\mathrm{CNOT}_{01}$, the $R_z$ rotations, and a trailing $\mathrm{CNOT}_{01}$; the Clifford+$T$ suffix ($\mathrm{CT}$ pairs, remaining CNOTs, $A',B'$) is unchanged.}}
    \label{fig:partial-decomposition}
\end{figure}

\black{Setting $U_2=U$, apply the two-CNOT decomposition of Sec.~\ref{sec:decomposition-three-cnots} to obtain four $\mathrm{CNOT}_{01}$ gates, four single-qubit unitaries $A,B,C,D$, and embedded rotations $R_x(\theta)$, $R_z(\phi)$, $R_z(-\psi)$, in a similar layout to Fig.~\ref{fig:two-qubit-decomposition} (top).}
\black{As in Sec.~\ref{sec:approximation-process-two}, magnitude-approximate the $R_x(\theta)$ and $R_z(\phi)$ that are sandwiched between $\mathrm{CNOT}_{01}$ gates and absorb the residual rotations into $A,B,C,D$; the result is the \emph{top} row of Fig.~\ref{fig:partial-decomposition}, which corresponds to the bottom line of Fig.~\ref{fig:two-qubit-decomposition} with primed blocks $A',B',C',D'$ and Clifford+$T$ cores ($\mathrm{CT}$).}
\black{From here the procedure differs from the full two-qubit synthesis in Sec.~\ref{sec:approximation-process-two}: only $C'$ and $D'$ are Euler--decomposed,}
\begin{align}
    \label{eq:euler-CD-prime}
    C' & = e^{i\phi_{C'}} e^{-i\phi_{1,C'} Z/2} e^{-i\theta_{C'} X/2} e^{-i\phi_{2,C'} Z/2}, \\
    D' & = e^{i\phi_{D'}} e^{-i\phi_{1,D'} Z/2} e^{-i\theta_{D'} X/2} e^{-i\phi_{2,D'} Z/2}.
\end{align}
\black{All factors in~\eqref{eq:euler-CD-prime} \emph{except} $e^{-i\phi_{2,C'} Z/2}$ and $e^{-i\phi_{2,D'} Z/2}$ are compiled to Clifford+$T$ (the $\mathrm{CT}$ boxes on each wire in the \emph{middle} row of Fig.~\ref{fig:partial-decomposition}); those two right Euler $Z$ rotations are \emph{not} synthesized and are kept as explicit phase degrees of freedom.
    The remaining single-qubit blocks $A'$ and $B'$ are synthesized with the usual single-qubit Clifford+$T$ procedure (\texttt{approximate\_one\_qubit\_unitary}).}
\black{Finally, collect the two $\mathrm{CNOT}_{01}$ gates that sandwich $R_z(-\psi)$ on the target, together with the deferred right Euler factors $e^{-i\phi_{2,C'} Z/2}$ and $e^{-i\phi_{2,D'} Z/2}$, into one unsynthesized diagonal bank:}
\begin{equation}
    \label{eq:delta-phase-bank}
    \begin{aligned}
        \Delta_{\text{phase}} := {} &
        \bigl(e^{-i\phi_{2,C'} Z/2} \otimes e^{-i\phi_{2,D'} Z/2}\bigr)\,
        \mathrm{CNOT}_{01}\,
        \bigl(I \otimes R_z(-\psi)\bigr)\,
        \mathrm{CNOT}_{01}.
    \end{aligned}
\end{equation}
\black{The \emph{bottom} row of Fig.~\ref{fig:partial-decomposition} depicts~\eqref{eq:delta-phase-bank} as a single two-qubit box $\Delta_{\text{phase}}$, followed by the same Clifford+$T$ tail as the middle row.}

\black{This partial decomposition strategy leaves the diagonal phase matrix $\Delta_{\text{phase}}$ undecomposed, allowing it to be absorbed into subsequent operations in multi-qubit synthesis.}
\black{This approach is particularly useful in recursive decomposition, where propagating phase corrections through the circuit structure can potentially reduce the overall gate count.}
\black{The detailed mechanism for absorbing this phase matrix in three-qubit and larger unitaries will be explained in Section~\ref{sec:multi-qubit-recursive}.}

\subsection{Multi-Qubit Recursive Decomposition}\label{sec:multi-qubit-recursive}

\black{Pseudocode~\ref{alg:multi-qubit-roadmap} gives the overall roadmap for multi-qubit Clifford+$T$ synthesis.
    The subsections below detail each component of the recursive decomposition:}
\begin{itemize}
    \item \black{Sec.~\ref{sec:block-zxz}: \textbf{Block ZXZ decomposition} separates one qubit from the remaining $n{-}1$ qubits, expressing $U$ as a product of multiplexors and a diagonal gate.}
    \item \black{Sec.~\ref{sec:demultiplexing}: \textbf{Demultiplexing} factors each multiplexor into two $(n{-}1)$-qubit unitaries and a multiplexed $R_z$ gate.}
    \item \black{Sec.~\ref{sec:multiplexed-rz-decomposition}: \textbf{Multiplexed $R_z$ gate decomposition} expands each multiplexed $R_z$ into a sequence of CNOT gates and single-qubit $R_z$ rotations via Gray code techniques.}
    \item \black{Sec.~\ref{sec:recursive-decomposition}: \textbf{Recursive decomposition algorithm} combines the above steps and recurses until all blocks are two-qubit or smaller.}
    \item \black{Sec.~\ref{sec:approx-decomposed-components}: \textbf{Approximation of decomposed components} approximates each component to Clifford+$T$ and exploits partial decomposition to reduce the $T$-count.}
\end{itemize}

\begin{algorithm}[tbp]
    \caption{Roadmap for multi-qubit Clifford+$T$ synthesis (Section~\ref{sec:multi-qubit}).}
    \label{alg:multi-qubit-roadmap}
    \begin{algorithmic}[1]
        \Require target unitary $U$ on $n$ qubits, tolerance $\epsilon$
        \Ensure Clifford+$T$ circuit and implemented unitary within $\epsilon$ of $U$ (diamond norm)
        \If{$n = 2$}
        \State Decompose $U$ into three CNOTs with single-qubit blocks; approximate embedded rotations and four $\mathrm{SU}(2)$ factors (Subsec.~\ref{sec:two-qubit}).
        \Else
        \State Apply block ZXZ decomposition to $U$ (Subsec.~\ref{sec:block-zxz}).
        \State Demultiplex multiplexor blocks into $(n{-}1)$-qubit unitaries and multiplexed $R_z$ gates (Subsec.~\ref{sec:demultiplexing}).
        \For{each $(n{-}1)$-qubit factor $V$ produced in these steps}
        \If{$V$ is a unitary on three or more qubits}
        \State Run this roadmap recursively on $V$ under the allocated diamond-norm budget for~$V$.
        \EndIf
        \EndFor
        \State Decompose each multiplexed $R_z$ gate into CNOTs and single-qubit $R_z$ gates (Gray code; Subsec.~\ref{sec:multiplexed-rz-decomposition}).
        \State Approximate all remaining two-qubit unitaries with their allocated shares of $\epsilon$; enable partial two-qubit decomposition so diagonal phases can be absorbed upstream (Subsec.~\ref{sec:approx-decomposed-components}).
        \EndIf
    \end{algorithmic}
\end{algorithm}

We use the following notions in the decomposition.
\begin{itemize}
    \item \textbf{Quantum multiplexor:}
          \begin{center}
              \begin{quantikz}
                  \lstick{} & \qw         & \octrl[style=octrlsquare]{1} & \qw \\
                  \lstick{} & \qwbundle{} & \gate{U}                     & \qw
              \end{quantikz}
              \quad $= U_1 \oplus U_2 = \begin{pmatrix} U_1 & 0 \\ 0 & U_2 \end{pmatrix}$.
          \end{center}
          The control selects $U_1$ when it is $\ket{0}$ and $U_2$ when it is $\ket{1}$. \black{Note that the notation is slightly different from controlled gates; $U$ indicates the $n$-qubit unitary instead of $(n-1)$-qubit unitary.
          }
    \item \textbf{\black{Multiplexed Pauli rotation gate:}}
          \begin{center}
              \begin{quantikz}
                  \lstick{} & \qwbundle{} & \octrl[style=octrlsquare]{1} & \qw \\
                  \lstick{} & \qw         & \gate{R_a}                    & \qw
              \end{quantikz}
          \end{center}
          The target qubit undergoes a rotation about Pauli axis $a$ whose angle is conditioned on the state of the control qubits.
\end{itemize}

\black{For unitaries acting on more than two qubits, the function \texttt{approximate\allowbreak\_multi\allowbreak\_qubit\allowbreak\_unitary} employs a recursive block decomposition strategy.}
This is our proposed extension to multi-qubit systems, building upon the existing two-qubit decomposition methods.

\subsubsection{Block ZXZ Decomposition}\label{sec:block-zxz}

The function \texttt{\_blockZXZ} decomposes an $n$-qubit unitary $U$ into a block structure~\cite{Krol2024beyond}:
\begin{equation}\label{eq:block-zxz}
    \begin{aligned}
        U & = \frac{1}{2} \begin{pmatrix} A_1 & 0 \\ 0 & A_2 \end{pmatrix} \begin{pmatrix} I+B & I-B \\ I-B & I+B \end{pmatrix} \begin{pmatrix} I & 0 \\ 0 & C \end{pmatrix} \\
          & = \frac{1}{2} \begin{pmatrix} A_1 & 0 \\ 0 & A_2 \end{pmatrix} (H \otimes I) \begin{pmatrix} I & 0 \\ 0 & B \end{pmatrix} (H \otimes I) \begin{pmatrix} I & 0 \\ 0 & C \end{pmatrix},
    \end{aligned}
\end{equation}
\black{where $A_1, A_2, B, C$ are $(n-1)$-qubit unitaries.
    We write $A = A_1 \oplus A_2$ for the multiplexor formed by the first factor.}
\black{$A_1$, $A_2$, $B$, and $C$ are determined by the SVD of the upper-left and upper-right blocks of $U$.}
This decomposition isolates one qubit from the rest, allowing recursive processing.
\black{The corresponding quantum circuit reads:}
\begin{equation}\label{eq:block-zxz-circuit}
    \begin{quantikz}
        \lstick{} & \qw         & \gate[wires=2]{U} & \qw \\
        \lstick{} & \qwbundle{} &                   & \qw
    \end{quantikz}
    \quad = \quad
    \begin{quantikz}
        \lstick{} & \qw         & \ctrl{1} & \gate{H} & \ctrl{1} & \gate{H} & \octrl[style=octrlsquare]{1} & \qw \\
        \lstick{} & \qwbundle{} & \gate{C}  & \qw      & \gate{B} & \qw      & \gate{A}  & \qw
    \end{quantikz}
\end{equation}
\black{Here $A$ denotes the multiplexor $A_1 \oplus A_2$; the controlled-$B$ gate implements $I \oplus B$; and the controlled-$C$ gate implements $I \oplus C$.}

\subsubsection{Demultiplexing}\label{sec:demultiplexing}

The function \texttt{\_demultiplex} is used to factor a pair of unitaries $M_1, M_2$ as~\cite{Krol2024beyond}:
\begin{equation}
    \begin{pmatrix} M_1 & 0 \\ 0 & M_2 \end{pmatrix}
    = \begin{pmatrix} V & 0 \\ 0 & V \end{pmatrix}
    \begin{pmatrix} \sqrt{D} & 0 \\ 0 & \sqrt{D^\dagger} \end{pmatrix}
    \begin{pmatrix} W & 0 \\ 0 & W \end{pmatrix},
\end{equation}
where $D$ is a diagonal matrix of eigenvalues of $M_1 M_2^\dagger$ and the block diagonal matrix $\sqrt{D} \oplus \sqrt{D^\dagger}$ corresponds to a multiplexed $R_z$ gate.
This allows controlled operations to be expressed in terms of $(n-1)$-qubit and diagonal operations.
For a multiplexor $M = M_1 \oplus M_2$, the decomposition is obtained by computing the eigenvalues of $M_1 M_2^\dagger$; these phases define the diagonal $D$ and determine the gates $V$, $W$, and the controlled $R_z$ in the circuit below.

In a quantum circuit, demultiplexing takes the following form:
\begin{equation}
    \begin{quantikz}
        \lstick{} & \qw         & \octrl[style=octrlsquare]{1} & \qw \\
        \lstick{} & \qwbundle{} & \gate{M}                     & \qw
    \end{quantikz}
    \quad = \quad
    \begin{quantikz}
        \lstick{} & \qw         & \qw      & \gate{R_z}                      & \qw      & \qw \\
        \lstick{} & \qwbundle{} & \gate{W} & \octrl[style=octrlsquare]{-1}   & \gate{V} & \qw
    \end{quantikz}
\end{equation}

\subsubsection{\black{Multiplexed $R_z$ Gate Decomposition}}\label{sec:multiplexed-rz-decomposition}

\begin{figure}[htbp]
    \centering
    \small
    \begin{quantikz}
        \lstick{} & \octrl[style=octrlsquare]{1} & \qw \\
        \lstick{} & \octrl[style=octrlsquare]{1} & \qw \\
        \lstick{} & \octrl[style=octrlsquare]{1} & \qw \\
        \lstick{} & \gate{R_z} & \qw
    \end{quantikz}
    \quad = \quad
    \begin{quantikz}[column sep=0.35em]
        \lstick{} & \qw                  & \qw      & \qw                  & \qw      & \qw                  & \qw      & \qw                  & \ctrl{3} & \qw                  & \qw      & \qw                  & \qw      & \qw                  & \qw      & \qw                  & \ctrl{3} & \qw \\
        \lstick{} & \qw                  & \qw      & \qw                  & \ctrl{2} & \qw                  & \qw      & \qw                  & \qw      & \qw                  & \qw      & \qw                  & \ctrl{2} & \qw                  & \qw      & \qw                  & \qw      & \qw \\
        \lstick{} & \qw                  & \ctrl{1} & \qw                  & \qw      & \qw                  & \ctrl{1} & \qw                  & \qw      & \qw                  & \ctrl{1} & \qw                  & \qw      & \qw                  & \ctrl{1} & \qw                  & \qw      & \qw \\
        \lstick{} & \gate{R_z(\theta_1)} & \targ{}  & \gate{R_z(\theta_2)} & \targ{}  & \gate{R_z(\theta_3)} & \targ{}  & \gate{R_z(\theta_4)} & \targ{}  & \gate{R_z(\theta_5)} & \targ{}  & \gate{R_z(\theta_6)} & \targ{}  & \gate{R_z(\theta_7)} & \targ{}  & \gate{R_z(\theta_8)} & \targ{}  & \qw
    \end{quantikz}
    \caption{\black{Example of multiplexed $R_z$ gate decomposition for a 4-qubit unitary: the control qubits (top) determine which $R_z$ rotation is applied to the target qubit (bottom).}}
    \label{fig:multiplexed-rz-decomposition}
\end{figure}
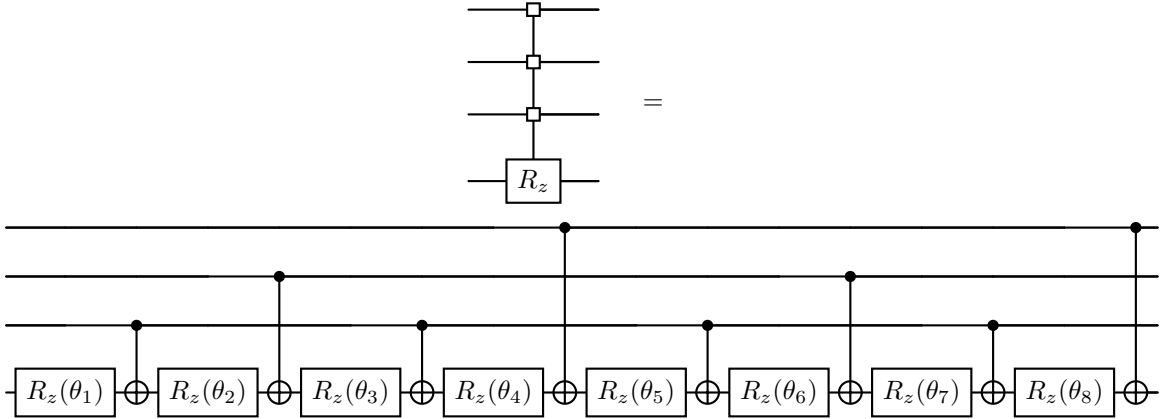

The function \texttt{\_decompose\_cz} decomposes \black{a multiplexed $R_z$ gate} into a sequence of CNOT gates and single-qubit $Z$ rotations~\cite{Krol2024beyond}.
\black{Concretely, a multiplexed $R_z$ gate on $n$ qubits ($n{-}1$ control qubits and one target) applies a different $R_z(\varphi_j)$ rotation to the target for each computational-basis state $\ket{j}$ of the controls, and therefore realizes a $2^{n}\times 2^{n}$ diagonal unitary.}
The decomposition uses Gray code ordering to minimize the number of gates, with the decomposition based on solving a linear system involving a matrix derived from Gray code patterns.
For example, a multiplexed $R_z$ gate on four qubits (three controls, one target) decomposes as shown in Fig.~\ref{fig:multiplexed-rz-decomposition}.

\black{In Fig.~\ref{fig:multiplexed-rz-decomposition}, the rotation angles $\alpha_1,\ldots,\alpha_{2^k}$ realised by the multiplexed gate are related to the decomposition parameters $\theta_1,\ldots,\theta_{2^k}$ by a linear \black{equation}~\cite{Krol2024beyond}:}
\begin{equation}\label{eq:multiplexed-rz-angles}
    M^{k}
        \begin{pmatrix} \theta_1 \\ \vdots \\ \theta_{2^k} \end{pmatrix}
        =
        \begin{pmatrix} \alpha_1 \\ \vdots \\ \alpha_{2^k} \end{pmatrix},
        \qquad
        M^{k}_{ij} = (-1)^{b_{i-1}\,\cdot\, g_{j-1}},
\end{equation}
\black{where $b_i$ is the standard binary representation of integer~$i$, $g_j$ is the binary representation of the $j$-th Gray code number, and the dot denotes the bitwise inner product~\cite{Krol2024beyond}.
    For instance, with $k=3$ controls, the rotation angle for control state $\ket{011}$ (i.e.\ the $i{=}4$ row of $M^3$, since $b_3 = 011$) evaluates to}
\begin{equation}\label{eq:multiplexed-rz-j3}
    \alpha_4 = +\theta_1 - \theta_2 + \theta_3 - \theta_4 - \theta_5 + \theta_6 - \theta_7 + \theta_8.
\end{equation}

\subsubsection{Recursive Decomposition Algorithm}\label{sec:recursive-decomposition}

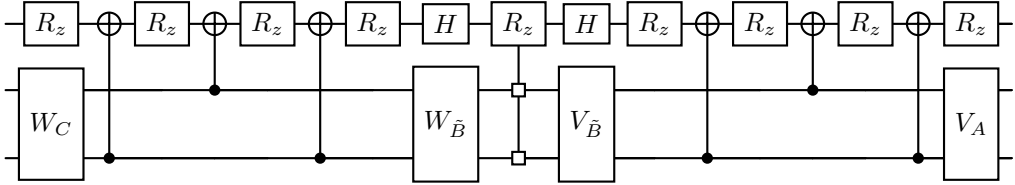
\begin{figure}[htbp]
    \centering
    \small
    \textbf{Step 1.} \black{Block ZXZ decomposition (Sec.~\ref{sec:block-zxz}): same form as~\eqref{eq:block-zxz-circuit}.}
    \begin{minipage}[t]{\linewidth}
        \centering
        \begin{quantikz}[column sep=0.5em]
            \lstick{} & \qw         & \ctrl{1} & \gate{H} & \ctrl{1} & \gate{H} & \octrl[style=octrlsquare]{1} & \qw \\
            \lstick{} & \qwbundle{} & \gate{C}  & \qw      & \gate{B} & \qw      & \gate{A}  & \qw
        \end{quantikz}
    \end{minipage}\\[0.6em]
    \textbf{Step 2.} \black{Demultiplex $A{=}A_1{\oplus}A_2$ and $I{\oplus}C$ (Sec.~\ref{sec:demultiplexing}).}
    \begin{minipage}[t]{\linewidth}
        \centering
        \begin{quantikz}[column sep=0.5em]
            \lstick{} & \qw         & \qw        & \gate{R_z}                    & \gate{H} & \ctrl{1} & \gate{H} & \gate{R_z}                    & \qw        & \qw \\
            \lstick{} & \qwbundle{} & \gate{W_C} & \octrl[style=octrlsquare]{-1} & \gate{V_C} & \gate{B} & \gate{W_A} & \octrl[style=octrlsquare]{-1} & \gate{V_A} & \qw
        \end{quantikz}
    \end{minipage}\\[0.6em]
    \textbf{Step 3.} \black{Decompose the two multiplexed $R_z$ gates (Sec.~\ref{sec:multiplexed-rz-decomposition}).}
    \begin{minipage}[t]{\linewidth}
        \centering
        \begin{quantikz}[column sep=0.35em, row sep=0.8em]
            \lstick{} & \gate{R_z} & \targ{} & \gate{R_z} & \targ{} & \gate{R_z} & \targ{} & \gate{R_z} & \targ{} & \gate{H} & \ctrl{1} & \gate{H} & \targ{} & \gate{R_z} & \targ{} & \gate{R_z} & \targ{} & \gate{R_z} & \targ{} & \gate{R_z} & \qw \\
            \lstick{} & \gate[wires=2]{W_C} & \qw & \qw & \ctrl{-1} & \qw & \qw & \qw & \ctrl{-1} & \gate[wires=2]{V_C} & \gate[wires=2]{B} & \gate[wires=2]{W_A} & \ctrl{-1} & \qw & \qw & \qw & \ctrl{-1} & \qw & \qw & \gate[wires=2]{V_A} & \qw \\
            \lstick{} &                     & \ctrl{-2} & \qw & \qw & \qw & \ctrl{-2} & \qw & \qw & & & & \qw & \qw & \ctrl{-2} & \qw & \qw & \qw & \ctrl{-2} &  &  & \qw
        \end{quantikz}
    \end{minipage}\\[0.6em]
    \textbf{Step 4.} \black{Merge $V_C$, $W_A$, controlled-$B$, and the adjacent two-qubit gates from Step~3 into $\tilde{B}$~\cite{Krol2024beyond}; demultiplex $\tilde{B}$ and decompose its multiplexed $R_z$.}
    \begin{minipage}[t]{\linewidth}
        \centering
        \begin{quantikz}[column sep=0.5em, row sep=0.8em]
            \lstick{} & \gate{R_z}          & \targ{}   & \gate{R_z} & \targ{}   & \gate{R_z} & \targ{}   & \gate{R_z} & \gate{H}                      & \gate{R_z}                    & \gate{H}                      & \gate{R_z} & \targ{}   & \gate{R_z} & \targ{}   & \gate{R_z} & \targ{}   & \gate{R_z}          & \qw \\
            \lstick{} & \gate[wires=2]{W_C} & \qw       & \qw        & \ctrl{-1} & \qw        & \qw       & \qw        & \gate[wires=2]{W_{\tilde{B}}} & \octrl[style=octrlsquare]{-1} & \gate[wires=2]{V_{\tilde{B}}} & \qw        & \qw       & \qw        & \ctrl{-1} & \qw        & \qw       & \gate[wires=2]{V_A} & \qw \\
            \lstick{} &                     & \ctrl{-2} & \qw        & \qw       & \qw        & \ctrl{-2} & \qw        & \qw                           & \octrl[style=octrlsquare]{-1} & \qw                           & \qw        & \ctrl{-2} & \qw        & \qw       & \qw        & \ctrl{-2} & \qw                 & \qw
        \end{quantikz}
    \end{minipage}
    \caption{\black{Step-by-step decomposition of a 3-qubit unitary.
    }
    }
    \label{fig:step-by-step-decomposition}
\end{figure}

\black{The subsections above define the three elementary tools---block ZXZ (Sec.~\ref{sec:block-zxz}), demultiplexing (Sec.~\ref{sec:demultiplexing}), and multiplexed $R_z$ decomposition (Sec.~\ref{sec:multiplexed-rz-decomposition}).
Fig.~\ref{fig:step-by-step-decomposition} shows how they are composed for $n{=}3$.
Block ZXZ (Step~1) is followed by demultiplexing the multiplexors $A{=}A_1{\oplus}A_2$ and $I{\oplus}C$ (Step~2), then by decomposing the two multiplexed $R_z$ gates that appear on the top wire (Step~3).
In Step~4, $V_C$, $W_A$, the controlled-$B$ gate, and the adjacent two-qubit gates introduced in Step~3 are merged into a single multiplexor $\tilde{B}$~\cite{Krol2024beyond}; $\tilde{B}$ is subsequently demultiplexed and its multiplexed $R_z$ gate is decomposed as well.}

The complete recursive decomposition algorithm \texttt{\_decompose\_recursively} works as follows:
\begin{enumerate}
    \item For a single or two-qubit unitary, return it directly or use the two-qubit decomposition.
    \item For $n > 2$ qubits:
          \begin{enumerate}
              \item Apply block ZXZ decomposition to separate the first qubit.
              \item \black{Demultiplex the resulting multiplexors, merge adjacent blocks to form $\tilde{B}$, and decompose all multiplexed $R_z$ gates (Steps~1--4 of Fig.~\ref{fig:step-by-step-decomposition}).}
              \item \black{Recursively decompose each $(n{-}1)$-qubit unitary by repeating this procedure until all remaining blocks are two-qubit unitaries.}
              \item \black{Decompose the multiplexed $R_z$ gates using Gray code techniques (Sec.~\ref{sec:multiplexed-rz-decomposition}).}
          \end{enumerate}
\end{enumerate}

\subsubsection{Approximation of Decomposed Components}\label{sec:approx-decomposed-components}

\black{Once the recursive decomposition (Sec.~\ref{sec:recursive-decomposition}) is complete, the circuit consists of three kinds of components:}
\begin{enumerate}
    \item \black{Single-qubit $R_z$ rotations --- approximated to Clifford+$T$ using \texttt{gridsynth\_circuit} (Sec.~\ref{sec:gridsynth}).}
    \item \black{CNOT and Hadamard gates --- kept as-is (they are already Clifford gates).}
    \item \black{Two-qubit unitaries from the recursive base case --- approximated using \texttt{approximate\allowbreak\_two\allowbreak\_qubit\allowbreak\_unitary} (Sec.~\ref{sec:two-qubit}).}
\end{enumerate}

\black{A key optimisation targets the two-qubit unitaries produced by the recursive base case.
    All intermediate two-qubit blocks (i.e.\ every block \emph{except} the last one encountered in the recursion) are approximated with \texttt{decompose\_partially=True}, using the partial decomposition described in Sec.~\ref{sec:partial-decomposition}.
    This leaves a residual diagonal phase matrix $\Delta_{\text{phase}}$~\eqref{eq:delta-phase-bank} unsynthesized alongside the Clifford+$T$ gates.}

\black{The crucial observation is that $\Delta_{\text{phase}}$ is diagonal in the computational basis, and multiplexed $R_z$ gates (Sec.~\ref{sec:multiplexed-rz-decomposition}) are also diagonal.}
Since both operations are diagonal, they commute:
\begin{equation}\label{eq:delta-commute-mrz}
    (\Delta_{\mathrm{MRz}} \oplus \Delta_{\mathrm{MRz}}^\dagger) \cdot (I^{\otimes (n-2)} \otimes \Delta_{\text{phase}}) = (I^{\otimes (n-2)} \otimes \Delta_{\text{phase}}) \cdot (\Delta_{\mathrm{MRz}} \oplus \Delta_{\mathrm{MRz}}^\dagger),
\end{equation}
\black{where $\Delta_{\mathrm{MRz}} \oplus \Delta_{\mathrm{MRz}}^\dagger$ denotes a multiplexed $R_z$ gate in block-diagonal form.
    This commutativity allows $\Delta_{\text{phase}}$ to pass through the adjacent multiplexed $R_z$ gate and be absorbed into the \emph{next} two-qubit block in the recursive circuit.}

\begin{figure}[p]
    \centering
    \footnotesize
    \textbf{0.} \black{Same as Step~4 of Fig.~\ref{fig:step-by-step-decomposition}.}
    \begin{minipage}[t]{\linewidth}
        \centering
        \begin{quantikz}[column sep=0.4em, row sep=0.6em]
            \lstick{} & \gate{R_z} & \targ{} & \gate{R_z} & \targ{} & \gate{R_z} & \targ{} & \gate{R_z} & \gate{H} & \gate{R_z} & \gate{H} & \gate{R_z} & \targ{} & \gate{R_z} & \targ{} & \gate{R_z} & \targ{} & \gate{R_z} & \qw \\
            \lstick{} & \gate[wires=2]{W_C} & \qw & \qw & \ctrl{-1} & \qw & \qw & \qw & \gate[wires=2]{W_{\tilde{B}}} & \octrl[style=octrlsquare]{-1} & \gate[wires=2]{V_{\tilde{B}}} & \qw & \qw & \qw & \ctrl{-1} & \qw & \qw & \gate[wires=2]{V_A} & \qw \\
            \lstick{} & & \ctrl{-2} & \qw & \qw & \qw & \ctrl{-2} & \qw & \qw & \octrl[style=octrlsquare]{-1} & \qw & \qw & \ctrl{-2} & \qw & \qw & \qw & \ctrl{-2} & \qw & \qw
        \end{quantikz}
    \end{minipage}\\[0.3em]
    \textbf{1.} \black{$V_A$ partially decomposed: $\Delta_{V_A}$ to the left, CT to the right.}
    \begin{minipage}[t]{\linewidth}
        \centering
        \begin{quantikz}[column sep=0.4em, row sep=0.6em]
            \lstick{} & \gate{R_z} & \targ{} & \gate{R_z} & \targ{} & \gate{R_z} & \targ{} & \gate{R_z} & \gate{H} & \gate{R_z} & \gate{H} & \gate{R_z} & \targ{} & \gate{R_z} & \targ{} & \gate{R_z} & \targ{} & \gate{R_z} & \qw & \qw \\
            \lstick{} & \gate[wires=2]{W_C} & \qw & \qw & \ctrl{-1} & \qw & \qw & \qw & \gate[wires=2]{W_{\tilde{B}}} & \octrl[style=octrlsquare]{-1} & \gate[wires=2]{V_{\tilde{B}}} & \qw & \qw & \qw & \ctrl{-1} & \qw & \qw & \gate[wires=2]{\Delta_{V_A}} & \gate[wires=2]{\mathrm{CT}} & \qw \\
            \lstick{} & & \ctrl{-2} & \qw & \qw & \qw & \ctrl{-2} & \qw & \qw & \octrl[style=octrlsquare]{-1} & \qw & \qw & \ctrl{-2} & \qw & \qw & \qw & \ctrl{-2} & & & \qw
        \end{quantikz}
    \end{minipage}\\[0.3em]
    \textbf{2.} \black{$\Delta_{V_A}$ commutes left past the right multiplexed $R_z$.}
    \begin{minipage}[t]{\linewidth}
        \centering
        \begin{quantikz}[column sep=0.4em, row sep=0.6em]
            \lstick{} & \gate{R_z} & \targ{} & \gate{R_z} & \targ{} & \gate{R_z} & \targ{} & \gate{R_z} & \gate{H} & \gate{R_z} & \gate{H} & \qw & \gate{R_z} & \targ{} & \gate{R_z} & \targ{} & \gate{R_z} & \targ{} & \gate{R_z} & \qw \\
            \lstick{} & \gate[wires=2]{W_C} & \qw & \qw & \ctrl{-1} & \qw & \qw & \qw & \gate[wires=2]{W_{\tilde{B}}} & \octrl[style=octrlsquare]{-1} & \gate[wires=2]{\Delta_{V_A}} & \gate[wires=2]{V_{\tilde{B}}} & \qw & \qw & \qw & \ctrl{-1} & \qw & \qw & \gate[wires=2]{\mathrm{CT}} & \qw \\
            \lstick{} & & \ctrl{-2} & \qw & \qw & \qw & \ctrl{-2} & \qw & \qw & \octrl[style=octrlsquare]{-1} & & \qw & \qw & \ctrl{-2} & \qw & \qw & \qw & \ctrl{-2} & & \qw
        \end{quantikz}
    \end{minipage}\\[0.3em]
    \textbf{3.} \black{$\Delta_{V_A}$ absorbed into $V_{\tilde{B}}$; result $V'_{\tilde{B}}$ partially decomposed.}
    \begin{minipage}[t]{\linewidth}
        \centering
        \begin{quantikz}[column sep=0.4em, row sep=0.6em]
            \lstick{} & \gate{R_z} & \targ{} & \gate{R_z} & \targ{} & \gate{R_z} & \targ{} & \gate{R_z} & \gate{H} & \gate{R_z} & \gate{H} & \qw & \gate{R_z} & \targ{} & \gate{R_z} & \targ{} & \gate{R_z} & \targ{} & \gate{R_z} & \qw \\
            \lstick{} & \gate[wires=2]{W_C} & \qw & \qw & \ctrl{-1} & \qw & \qw & \qw & \gate[wires=2]{W_{\tilde{B}}} & \octrl[style=octrlsquare]{-1} & \gate[wires=2]{\Delta_{V'_{\tilde{B}}}} & \gate[wires=2]{\mathrm{CT}} & \qw & \qw & \qw & \ctrl{-1} & \qw & \qw & \gate[wires=2]{\mathrm{CT}} & \qw \\
            \lstick{} & & \ctrl{-2} & \qw & \qw & \qw & \ctrl{-2} & \qw & \qw & \octrl[style=octrlsquare]{-1} & & \qw & \qw & \ctrl{-2} & \qw & \qw & \qw & \ctrl{-2} & & \qw
        \end{quantikz}
    \end{minipage}\\[0.3em]
    \textbf{4.} \black{$\Delta_{V'_{\tilde{B}}}$ commutes left past the central multiplexed $R_z$; absorbed into $W_{\tilde{B}}$.}
    \begin{minipage}[t]{\linewidth}
        \centering
        \begin{quantikz}[column sep=0.4em, row sep=0.6em]
            \lstick{} & \gate{R_z} & \targ{} & \gate{R_z} & \targ{} & \gate{R_z} & \targ{} & \gate{R_z} & \gate{H} & \gate{R_z} & \gate{H} & \gate{R_z} & \targ{} & \gate{R_z} & \targ{} & \gate{R_z} & \targ{} & \gate{R_z} & \qw \\
            \lstick{} & \gate[wires=2]{W_C} & \qw & \qw & \ctrl{-1} & \qw & \qw & \qw & \gate[wires=2]{W'_{\tilde{B}}} & \octrl[style=octrlsquare]{-1} & \gate[wires=2]{\mathrm{CT}} & \qw & \qw & \qw & \ctrl{-1} & \qw & \qw & \gate[wires=2]{\mathrm{CT}} & \qw \\
            \lstick{} & & \ctrl{-2} & \qw & \qw & \qw & \ctrl{-2} & \qw & \qw & \octrl[style=octrlsquare]{-1} & \qw & \qw & \ctrl{-2} & \qw & \qw & \qw & \ctrl{-2} & & \qw
        \end{quantikz}
    \end{minipage}\\[0.3em]
    \textbf{5.} \black{$W'_{\tilde{B}}$ partially decomposed: $\Delta_{W'_{\tilde{B}}}$ to the left; commutes left past the left multiplexed $R_z$.}
    \begin{minipage}[t]{\linewidth}
        \centering
        \begin{quantikz}[column sep=0.4em, row sep=0.6em]
            \lstick{} & \gate{R_z} & \targ{} & \gate{R_z} & \targ{} & \gate{R_z} & \targ{} & \gate{R_z} & \qw & \gate{H} & \gate{R_z} & \gate{H} & \gate{R_z} & \targ{} & \gate{R_z} & \targ{} & \gate{R_z} & \targ{} & \gate{R_z} & \qw \\
            \lstick{} & \gate[wires=2]{W_C} & \qw & \qw & \ctrl{-1} & \qw & \qw & \gate[wires=2]{\Delta_{W'_{\tilde{B}}}} & \gate[wires=2]{\mathrm{CT}} & \octrl[style=octrlsquare]{-1} & \gate[wires=2]{\mathrm{CT}} & \qw & \qw & \qw & \ctrl{-1} & \qw & \qw & \gate[wires=2]{\mathrm{CT}} & \qw \\
            \lstick{} & & \ctrl{-2} & \qw & \qw & \qw & \ctrl{-2} & & \qw & \octrl[style=octrlsquare]{-1} & \qw & \qw & \ctrl{-2} & \qw & \qw & \qw & \ctrl{-2} & & \qw
        \end{quantikz}
    \end{minipage}\\[0.3em]
    \textbf{6.} \black{$W_C$ fully decomposed (absorbs $\Delta_{W'_{\tilde{B}}}$; no residual $\Delta$).}
    \begin{minipage}[t]{\linewidth}
        \centering
        \begin{quantikz}[column sep=0.4em, row sep=0.6em]
            \lstick{} & \gate{R_z} & \targ{} & \gate{R_z} & \targ{} & \gate{R_z} & \targ{} & \gate{R_z} & \gate{H} & \gate{R_z} & \gate{H} & \gate{R_z} & \targ{} & \gate{R_z} & \targ{} & \gate{R_z} & \targ{} & \gate{R_z} & \qw \\
            \lstick{} & \gate[wires=2]{\mathrm{CT}} & \qw & \qw & \ctrl{-1} & \qw & \qw & \qw & \gate[wires=2]{\mathrm{CT}} & \octrl[style=octrlsquare]{-1} & \gate[wires=2]{\mathrm{CT}} & \qw & \qw & \qw & \ctrl{-1} & \qw & \qw & \gate[wires=2]{\mathrm{CT}} & \qw \\
            \lstick{} & & \ctrl{-2} & \qw & \qw & \qw & \ctrl{-2} & \qw & \qw & \octrl[style=octrlsquare]{-1} & \qw & \qw & \ctrl{-2} & \qw & \qw & \qw & \ctrl{-2} & & \qw
        \end{quantikz}
    \end{minipage}
    \caption{\black{Recursive absorption of residual diagonal phase matrix for $n=3$: (0)~recursive circuit; (1)~$V_A$ partially decomposed to CT and $\Delta_{V_A}$; (2)~$\Delta_{V_A}$ commutes left past the right multiplexed $R_z$; (3)~absorbed into $V_{\tilde{B}}$, then $V'_{\tilde{B}}$ partially decomposed to CT and $\Delta_{V'_{\tilde{B}}}$; (4)~$\Delta_{V'_{\tilde{B}}}$ commutes left, absorbed into $W_{\tilde{B}} \to W'_{\tilde{B}}$; (5)~$W'_{\tilde{B}}$ partially decomposed, $\Delta_{W'_{\tilde{B}}}$ commutes left; (6)~$W_C$ fully decomposed (absorbs all accumulated $\Delta$; no residual). CT denotes a two-qubit block approximated by Clifford+$T$ gates.}}
    \label{fig:delta-phase-absorption}
\end{figure}
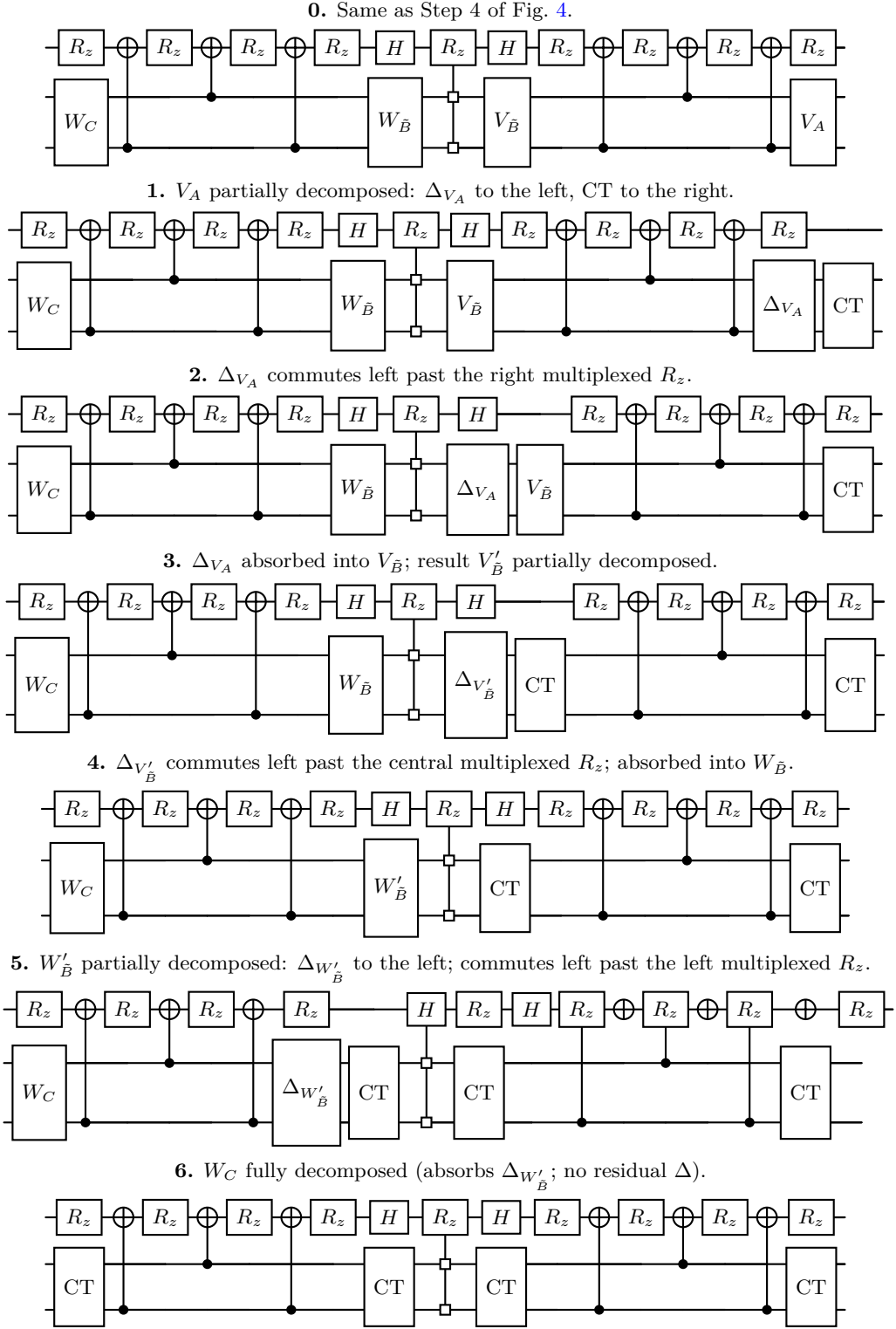

\black{The absorption proceeds as follows (Fig.~\ref{fig:delta-phase-absorption} illustrates the $n{=}3$ case):
    starting from the rightmost two-qubit block $V_A$ (step~1), partial decomposition produces Clifford+$T$ gates and a residual $\Delta_{V_A}$ to the left;
    the residual commutes leftward past the neighbouring multiplexed $R_z$ gate and is absorbed into the next two-qubit block;
    this process repeats until the \emph{first} two-qubit block ($W_C$ in the figure), which is fully decomposed so that no residual $\Delta$ remains.
    By avoiding the synthesis of intermediate $\Delta_{\text{phase}}$ matrices, this strategy significantly reduces the overall $T$-count, with the reduction becoming more pronounced as the number of qubits increases.}

After approximation, all components are combined to form the final Clifford+$T$ circuit.
The circuit has a hierarchical structure with controlled operations and recursively decomposed subcircuits, where each component is approximated using the appropriate synthesis method.

\black{The function \texttt{approximate\_multi\_qubit\_unitary} (source code~\ref{lst:multi-qubit-api}) is the top-level entry point for multi-qubit synthesis.
    It applies the recursive decomposition of Sec.~\ref{sec:recursive-decomposition} level by level, distributes the target approximation error $\epsilon$ across the recursion depth, and approximates each leaf block using the single- and two-qubit synthesizers (Sec.~\ref{sec:gridsynth} and~\ref{sec:two-qubit}).}
For multi-qubit unitaries with four or more qubits, it is not generally guaranteed that exact Clifford+$T$ decomposition can be achieved \black{when global phase is fixed}.
\black{For example, $e^{i\pi/8}I_{16}$ is in $SU(2^4)$, yet its entries $e^{i\pi/8}$ are not in $\mathbb{D}[\omega]$, which obstructs exact Clifford+$T$ synthesis under this requirement.}
Therefore, our synthesis framework performs approximation up to an overall global phase in these cases.
That is, the output circuit approximates the input unitary modulo a global phase factor, which is physically unobservable and does not affect any measurement outcomes or quantum computations.

\subsubsection{$T$-count Analysis for Multi-Qubit Synthesis}

The $T$-count for multi-qubit synthesis can be analyzed by counting the number of magnitude approximations and GridSynth approximations required in the recursive decomposition.
Let $M_n$ denote the number of magnitude approximations and $D_n$ denote the number of GridSynth approximations for an $n$-qubit unitary.

For small cases, we have $M_1 = 1$, $M_2 = 6$, $D_1 = 2$, and $D_2 = 9$.
For $n \geq 3$, the recursive structure yields:
\begin{align}
    M_n & = 6 \cdot 4^{n-2},                       \\
    D_n & = 3 \cdot 4^{n-1} - 3 \cdot 2^{n-1} + 3.
\end{align}

Since each magnitude approximation contributes $O(\log(1/\epsilon))$ $T$ gates and each GridSynth approximation contributes $O(3\log(1/\epsilon))$ $T$ gates, the total $T$-count is:
\begin{align}
    T_{\text{multi-qubit}} & = M_n \cdot \black{\log_2(1/\epsilon)} + D_n \cdot \black{3\log_2(1/\epsilon)} \black{+ o(\log(1/\epsilon))} \\
                           & = (M_n + 3D_n) \cdot \black{\log_2(1/\epsilon) + o(\log(1/\epsilon))}.
\end{align}
Substituting the expressions for $M_n$ and $D_n$:
\begin{align}
    M_n + 3D_n & = 6 \cdot 4^{n-2} + 3(3 \cdot 4^{n-1} - 3 \cdot 2^{n-1} + 3) \\
               & = 6 \cdot 4^{n-2} + 9 \cdot 4^{n-1} - 9 \cdot 2^{n-1} + 9    \\
               & = \frac{21}{8} \cdot 4^n - \frac{9}{2} \cdot 2^n + 9.
\end{align}
Therefore, the total $T$-count for $n$-qubit synthesis is:
\begin{equation}
    T_{\text{multi-qubit}} = \black{\left(\frac{21}{8} \cdot 4^n - \frac{9}{2} \cdot 2^n + 9\right)\log_2(1/\epsilon) + o(\log(1/\epsilon))}.
\end{equation}
This shows that the $T$-count scales exponentially with the number of qubits, but with a logarithmic dependence on the error tolerance.
The partial decomposition strategy helps reduce the constant factors, but the exponential scaling with $n$ is inherent to the recursive decomposition approach.

\section{Mixed Synthesis}\label{sec:mixed}

\black{Mixed unitary synthesis~\cite{Campbell2017shorter,Hastings2017turning,akibue2024probabilistic,Kliuchnikov2023shorterquantum,Yoshioka2025error} approximates a target channel by a classical mixture of Clifford+$T$ circuits, rather than by a single circuit.
    Sec.~\ref{sec:mixed-channels} defines the channel framework and the quadratic-error guarantee; Sec.~\ref{sec:mixed-ptm}--\ref{sec:mixed-lp} formulate the PTM-based LP that finds the optimal mixing probabilities; Sec.~\ref{sec:mixed-choi} evaluates diamond-norm error via Choi states; and Sec.~\ref{sec:mixed-impl} summarises the workflow.
    Source code~\ref{lst:mixed-api} shows a minimal call.}
\begin{lstlisting}[caption={Mixed synthesis: \texttt{mixed\_synthesis\_parallel}},label={lst:mixed-api}]
circuit_list, eu_np_list, probs, u_choi, u_choi_opt = mixed_synthesis_parallel(
    unitary=U,
    num_qubits=2,
    eps=1e-5,
    M=32,
    seed=123,
)
print(len(circuit_list), probs.sum(), probs.max())
\end{lstlisting}
\noindent\textbf{Output (illustrative).}
\begin{lstlisting}[style=codeoutput]
32 1.0 0.06832472415856752
\end{lstlisting}

\subsection{Quantum Channels and Probabilistic Mixtures}\label{sec:mixed-channels}

We aim to implement a target unitary $\hat{U}$ with its channel representation given by $\hat{\mathcal{U}}$, where $\hat{\mathcal{U}}(\rho) = \hat{U} \rho \hat{U}^\dagger$ for any density matrix $\rho$.
If we apply unitary $U_j$ with probability $p_j$ (where $p_j \geq 0$ and $\sum_j p_j = 1$), the resulting channel is:
\begin{equation}
    \tilde{\mathcal{U}}(\rho) = \sum_{j=1}^M p_j U_j \rho U_j^\dagger = \sum_{j=1}^M p_j \mathcal{U}_j(\rho),
\end{equation}
where $\mathcal{U}_j$ denotes the channel representation of $U_j$.

The key idea behind mixed synthesis is to approximate $\hat{\mathcal{U}}$ by generating a set of slightly shifted unitaries $\{U_j\}$ that are $\epsilon$-close approximations of $\hat{U}$ in terms of diamond \black{norm}: 
\begin{equation}
    \black{d_{\diamond}(\hat{\mathcal{U}}, \mathcal{U}_j) = \norm{\hat{\mathcal{U}} - \mathcal{U}_j}_\diamond \leq \epsilon.}
\end{equation}
There is a remarkable advantage of mixed synthesis when the unitaries $\{U_j\}$ constitute an $\epsilon$-net of the target unitary channel $\hat{\mathcal{U}}$ in terms of the diamond norm: it allows quadratic suppression of the error compared to unitary synthesis.
By appropriately generating a set of synthesized unitaries $\{U_j\}$ and choosing the optimal probability distribution over them, the diamond \black{norm} error can be reduced from $\epsilon$ to $\epsilon^2\black{/2}$, as follows.

\begin{lemma}[\cite{akibue2024probabilistic, Yoshioka2025error}]\label{lem:mixed-synth}
    Let $\hat{\mathcal{U}}$ be a target unitary channel with $\{\mathcal{U}_j\}$ constituting its $\epsilon$-net in terms of the diamond \black{norm error}.
    Then, one can find the optimal weight $\{p_j\}$ such that the diamond \black{norm error}
     is bounded as
    \begin{equation}
        \black{d_{\diamond}(\hat{\mathcal{U}}, \sum_j p_j \mathcal{U}_j) = \norm{\hat{\mathcal{U}} - \sum_j p_j \mathcal{U}_j}_\diamond \leq \frac{1}{2} \epsilon^2.}
        \label{eq:mixed-synth-bound}
    \end{equation}
\end{lemma}

\black{Note that the explicit constant in~\eqref{eq:mixed-synth-bound}  differs slightly from the bounds as stated in~\cite{akibue2024probabilistic, Yoshioka2025error}, because \black{we use the diamond norm error for $d_{\diamond}$ instead of the half-diamond norm error employed in the existing works.}
}

This quadratic error suppression makes mixed synthesis particularly useful when high-fidelity approximations are required, as it can achieve the same error tolerance with fewer $T$ gates or a larger error tolerance for the individual unitaries in the mixture.
However, directly minimizing the diamond norm $\norm{\hat{\mathcal{U}} - \sum_j p_j \mathcal{U}_j}_\diamond$ requires solving multiple semidefinite programming problems, which is computationally very inefficient, especially for multi-qubit systems.

\black{The $M$ perturbed unitaries $U_i = \hat{U}\exp(i\epsilon H_i)$ are generated by sampling random Hermitian operators $H_i$ (Appendix~\ref{sec:perturbation-appendix}), and each $U_i$ is then approximated using the multi-qubit synthesis of Section~\ref{sec:multi-qubit}.
    Heuristically, because an $n$-qubit unitary depends on $\sim\!4^n$ real parameters, we expect $M \gg 4^n$ to yield a sufficiently rich $\epsilon$-net for the mixture step.}

\subsection{Pauli Transfer Matrix Representation and Lightweight Cost Function}\label{sec:mixed-ptm}

To avoid the computational inefficiency of repeatedly solving semidefinite programming problems for diamond norm minimization, we use a lightweight cost function based on the Pauli Transfer Matrix (PTM) representation.
For an $n$-qubit system with Hilbert space dimension $d = 2^n$, the PTM is a $4^n \times 4^n$ matrix constructed from a Pauli operator basis $P_0, \ldots, P_{4^n-1} \in \{I, X, Y, Z\}^{\otimes n}$.
Each element of the PTM for a channel $\mathcal{U}$ is defined as:
\begin{equation}
    M_{ji} = \frac{1}{d} \mathrm{Tr}[P_j \mathcal{U}(P_i)],
\end{equation}
where $d = 2^n$ is the Hilbert space dimension.

The PTM representation allows us to express the mixed channel as a linear combination of individual channel PTMs:
\begin{equation}
    \mathrm{PTM}(\tilde{\mathcal{U}})_{ji} = \sum_{k=1}^M p_k \mathrm{PTM}(\mathcal{U}_k)_{ji},
\end{equation}
for all $i, j \in \{0, \ldots, 4^n-1\}$.
The approximation constraint is then to minimize the difference between the mixed PTM and the target PTM.

Instead of minimizing the diamond norm directly, we minimize the sum of absolute errors of each PTM element:
\begin{equation}
    \sum_{i,j} \left| \mathrm{PTM}(\hat{\mathcal{U}})_{ji} - \sum_{k=1}^M p_k \mathrm{PTM}(\mathcal{U}_k)_{ji} \right|,
\end{equation}
which provides a computationally tractable approximation to the original optimization problem.
This approach transforms the problem into a linear programming problem, which can be solved much more efficiently than semidefinite programming.

\subsection{Linear Programming Optimization}\label{sec:mixed-lp}

The function \texttt{compute\_optimal\_mixing\_probabilities} constructs a linear system from the PTM representations to find the optimal mixing probabilities.
Let $A$ be a matrix where each column corresponds to the vectorized PTM representation of a perturbed unitary $\mathcal{U}_i$:
\begin{equation}
    A = \begin{pmatrix} \mathrm{vec}(\mathrm{PTM}(\mathcal{U}_1)) & \cdots & \mathrm{vec}(\mathrm{PTM}(\mathcal{U}_M)) \end{pmatrix}^T,
\end{equation}
and let $b$ be the vectorized PTM representation of the target unitary:
\begin{equation}
    b = \mathrm{vec}(\mathrm{PTM}(\hat{\mathcal{U}})).
\end{equation}
Since the PTM is a $4^n \times 4^n$ matrix, the vectorized representation has dimension $4^{2n}$.
The optimization problem is to minimize the $L_1$ norm of the difference between the mixed PTM and the target PTM:
\begin{equation}
    \min_p \quad \norm{A p - b}_1 \quad \text{such that} \quad \sum_j p_j = 1, \quad p_j \geq 0.
\end{equation}

The function \texttt{solve\_LP} solves this linear programming problem by introducing slack variables $r$ to represent the absolute deviations:
\begin{align}
    \min_{p, r} & \quad \sum_i r_i                        \\
    \text{s.t.} & \quad \sum_i p_i \cdot \epsilon = 1,    \\
                & \quad p_i \geq 0,                       \\
                & \quad r_i \geq 0,                       \\
                & \quad A p - b / \epsilon \leq r / s,    \\
                & \quad -(A p - b / \epsilon) \leq r / s,
\end{align}
where $\epsilon$ is the error tolerance parameter and $s$ is a scaling factor (typically $s = 10^{-2} / \epsilon$).
The slack variables $r$ measure the deviation from the exact PTM matching, and the constraints ensure that $|A p - b / \epsilon| \leq r / s$ element-wise.
After solving, the probabilities are normalized so that $\sum_i p_i = 1$.

The function \texttt{solve\_LP} implements this linear program using convex optimization solvers (GUROBI, SCS, ECOS, or OSQP), automatically selecting an available solver.
We recommend using GUROBI when available, as it provides reliable solutions for this optimization problem.
This approach provides a computationally efficient alternative to directly minimizing the diamond norm, which would require solving multiple semidefinite programming problems.

\subsection{Choi Representation and Diamond Norm}\label{sec:mixed-choi}

For error analysis, we use the Choi representation of quantum channels.
For an $n$-qubit system with Hilbert space dimension $d = 2^n$, the Choi matrix of a unitary channel $\mathcal{U}$ is defined as:
\begin{equation}
    \mathcal{J}(\mathcal{U}) = (\mathcal{I}_d \otimes \mathcal{U})(\ket{\Omega}\bra{\Omega}),
\end{equation}
where $\mathcal{I}_d$ is the identity operation on a $d$-dimensional system, and $\ket{\Omega} = \sum_{i=1}^d \ket{i} \otimes \ket{i}$ is the unnormalized maximally entangled state on a $2n$-qubit system.
The Choi matrix provides a unique representation of the quantum channel.

The error between the target unitary channel and the mixed approximation is evaluated using the diamond norm:
\begin{equation}
    \norm{\hat{\mathcal{U}} - \tilde{\mathcal{U}}}_\diamond = \max_{\rho \in \mathbb{D}(\mathcal{H}_1 \otimes \mathcal{H}_2)} \norm{(\hat{\mathcal{U}} \otimes \mathcal{I}_d)(\rho) - (\tilde{\mathcal{U}} \otimes \mathcal{I}_d)(\rho)}_1,
\end{equation}
where $\norm{X}_1 = \mathrm{Tr}[\sqrt{X^\dagger X}]$ is the trace norm, $\mathcal{I}_d$ is the identity operation on a $d$-dimensional ancilla system, and $\mathbb{D}(\mathcal{H})$ denotes the set of physical density matrices on Hilbert space $\mathcal{H}$.
The diamond norm measures the worst-case error over all input states, including entangled states with an ancilla.

The function \texttt{diamond\_norm\_choi} computes the diamond norm using the Choi representation via semidefinite programming.
Given the Choi matrices $\mathcal{J}(\hat{\mathcal{U}})$ and $\mathcal{J}(\tilde{\mathcal{U}})$ of the target and mixed channels, the difference Choi matrix is $\mathcal{J}_\Delta = \mathcal{J}(\hat{\mathcal{U}}) - \mathcal{J}(\tilde{\mathcal{U}})$.
The diamond norm is then computed by solving the following semidefinite programming problem:
\begin{align}
    \min        & \quad t                                                    \\
    \text{s.t.} & \quad X \succeq 0, \quad X - \mathcal{J}_\Delta \succeq 0, \\
                & \quad Y_{ij} = \sum_{k=0}^{d-1} X_{i d + k, j d + k},      \\
                & \quad t I_d - Y \succeq 0, \quad t I_d + Y \succeq 0,
\end{align}
where $X$ is an $n \times n$ Hermitian matrix variable (with $n = d^2$), $Y$ is the partial trace of $X$ over the second subsystem, and $t$ is a scalar variable.
The diamond norm is then given by $2t$ (after appropriate scaling).
This semidefinite programming formulation allows for efficient computation of the diamond norm using convex optimization solvers.

\subsection{Implementation and Usage}\label{sec:mixed-impl}

The framework provides two implementations for mixed synthesis:
\begin{itemize}
    \item \texttt{mixed\_synthesis\_parallel}: Uses multiprocessing to approximate multiple perturbed unitaries in parallel, significantly reducing computation time.
    \item \texttt{mixed\_synthesis\_sequential}: Processes unitaries sequentially, useful when parallel processing is not available or when debugging.
\end{itemize}

Both functions follow the same overall workflow:
\begin{enumerate}
    \item \black{Generate perturbed unitaries $U_i = \hat{U} \exp(i \epsilon H_i)$ from random Hermitian operators (Appendix~\ref{sec:perturbation-appendix}).}
    \item Approximate each perturbed unitary using the multi-qubit synthesis algorithm described in Section~\ref{sec:multi-qubit}.
          The function \texttt{approximate\_unitary\_task} encapsulates this step, returning the synthesized circuit, approximated unitary matrix, PTM representation, and Choi representation for each perturbed unitary.
    \item Convert approximated unitaries to PTM and Choi representations.
    \item Solve the linear programming problem to find optimal mixing probabilities.
    \item Return the list of circuits, their corresponding mixing probabilities, and error metrics.
\end{enumerate}

\section{Numerical Experiments}\label{sec:experiments}

We conducted numerical experiments to validate the theoretical predictions and demonstrate the effectiveness of our synthesis framework.
The experiments were performed using the \texttt{pygridsynth} library implementation.

\subsection{Unitary synthesis benchmarks}

\begin{figure}[htbp]
    \centering
    \small
    \begin{minipage}[c]{0.47\linewidth}
        \vspace{0pt}
        \centering
        \includegraphics[width=\linewidth]{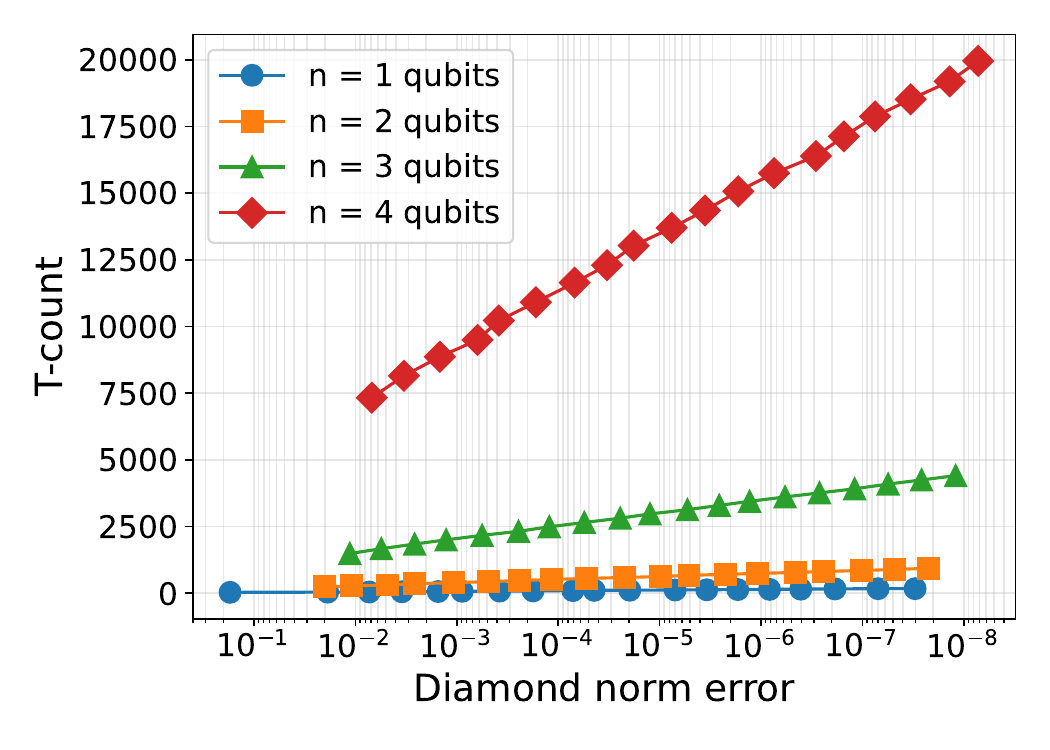}\\[-0.2em]
        {\footnotesize (a) All $n$.}
    \end{minipage}\hfill
    \begin{minipage}[c]{0.47\linewidth}
        \vspace{0pt}
        \centering
        \begin{minipage}[t]{0.48\linewidth}
            \vspace{0pt}
            \centering
            \includegraphics[width=\linewidth]{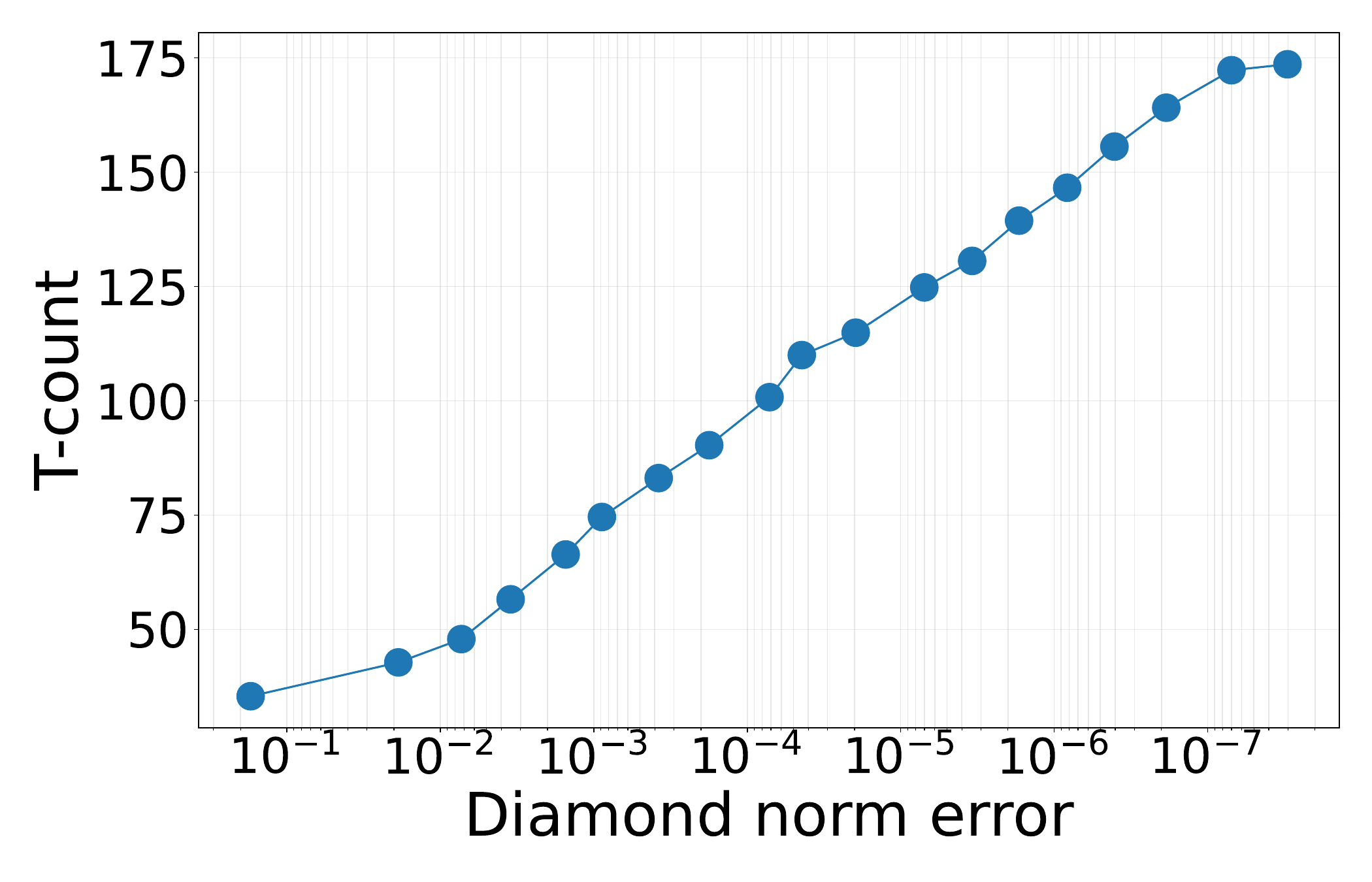}\\[-0.25em]
            {\footnotesize (b) $n{=}1$}
        \end{minipage}\hfill
        \begin{minipage}[t]{0.48\linewidth}
            \vspace{0pt}
            \centering
            \includegraphics[width=\linewidth]{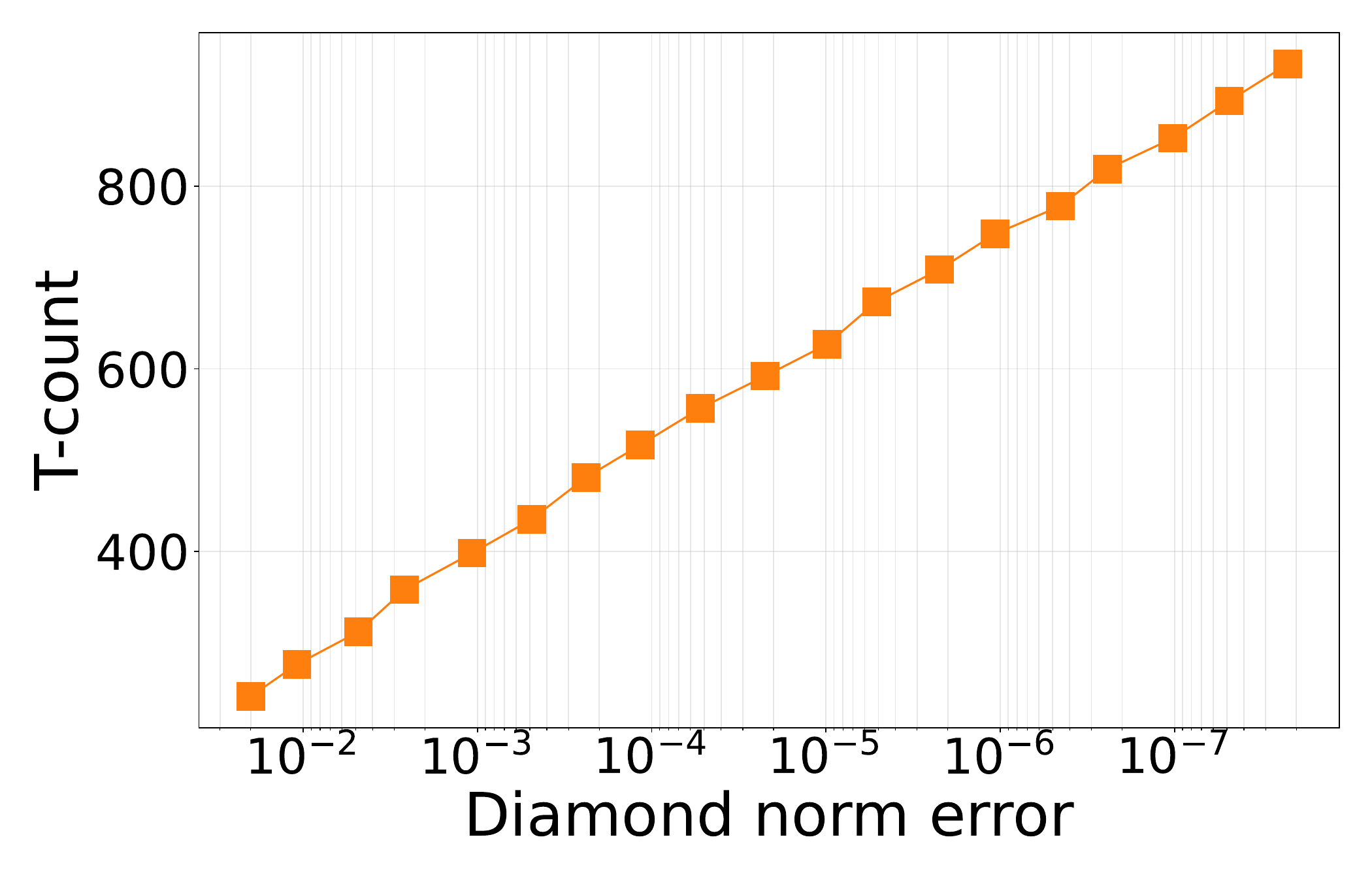}\\[-0.25em]
            {\footnotesize (c) $n{=}2$}
        \end{minipage}\\[0.35em]
        \begin{minipage}[t]{0.48\linewidth}
            \vspace{0pt}
            \centering
            \includegraphics[width=\linewidth]{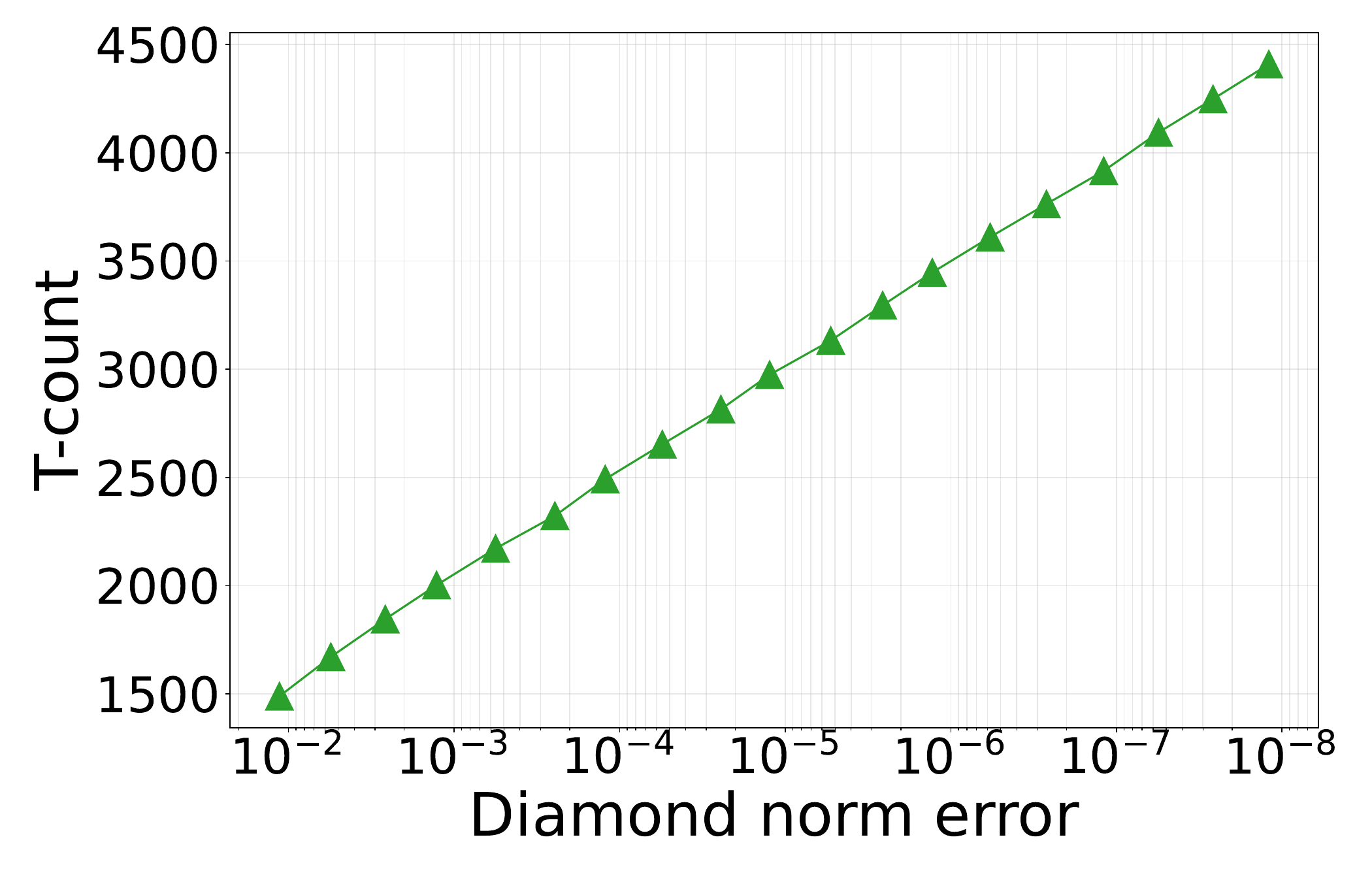}\\[-0.25em]
            {\footnotesize (d) $n{=}3$}
        \end{minipage}\hfill
        \begin{minipage}[t]{0.48\linewidth}
            \vspace{0pt}
            \centering
            \includegraphics[width=\linewidth]{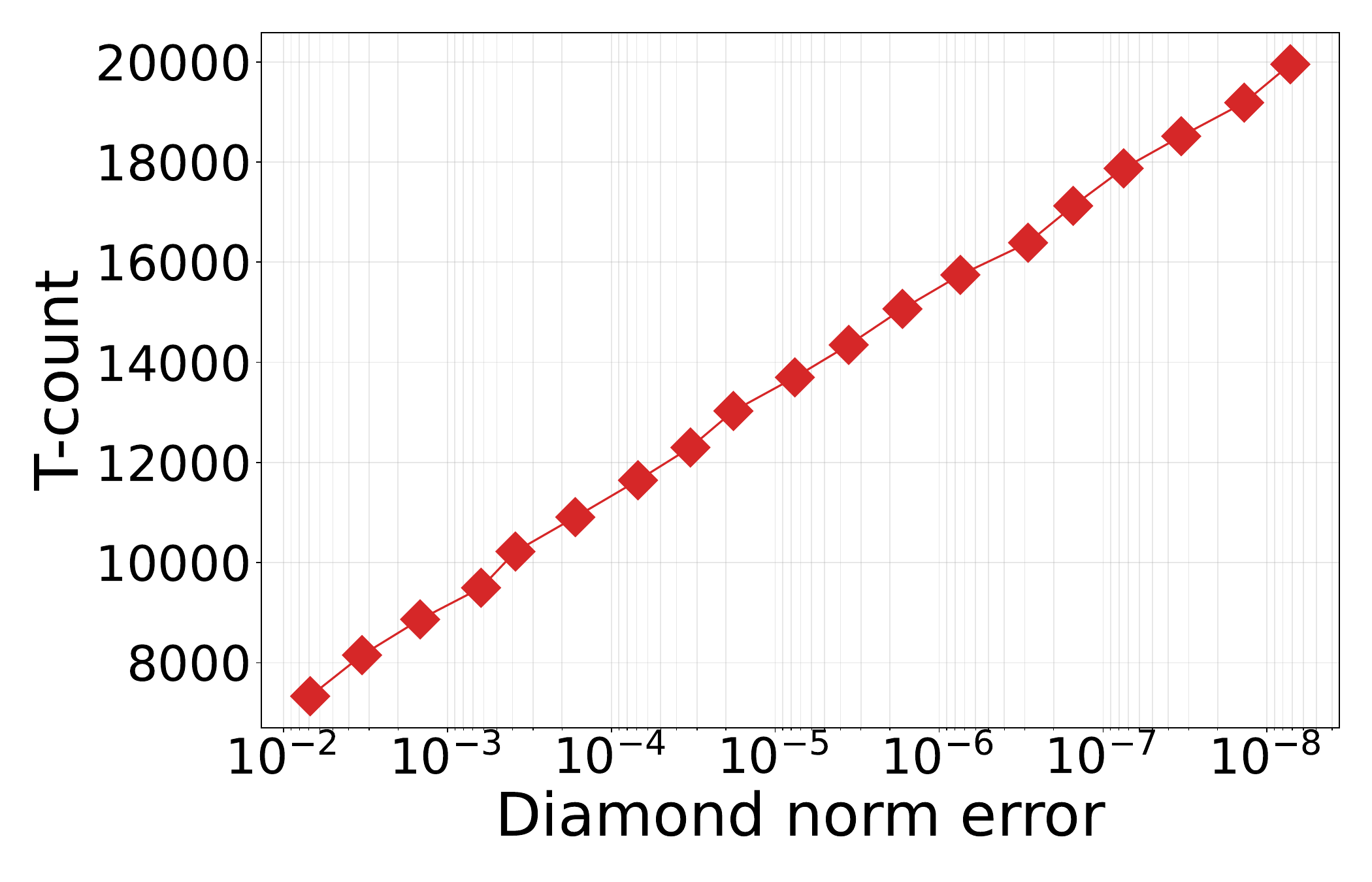}\\[-0.25em]
            {\footnotesize (e) $n{=}4$}
        \end{minipage}
    \end{minipage}
    \caption{Unitary synthesis benchmarks: each point corresponds to one synthesis run at a chosen tolerance $\epsilon$.
        Vertical axis: $T$-count of the output Clifford+$T$ circuit; horizontal axis: diamond-norm error between the target unitary channel and the implemented channel.
        (a)~All $n$ superposed.
        (b)--(e)~Same quantities for $n=1,2,3,4$ only.}
    \label{fig:dnorm-tcount-benchmark}
\end{figure}

We benchmarked Clifford+$T$ unitary synthesis (Sections~\ref{sec:single-qubit} and~\ref{sec:multi-qubit}) by sweeping the synthesis error tolerance $\epsilon$ and recording, for each resulting circuit, the diamond-norm distance between the target unitary channel and the synthesized channel and the total $T$-count.
Figure~\ref{fig:dnorm-tcount-benchmark} reports these tradeoffs.
Panel~(a) overlays all qubit numbers $n$ in one plot; panels~(b)--(e) use the same horizontal and vertical axes but show only $n=1,2,3,4$ so that slopes and point spacing can be read without overlap.

Table~\ref{tab:unitary-synth-time} reports the mean wall-clock time for one full unitary synthesis call in the same benchmark, for each~$n$ at $\epsilon=10^{-3}$ and $\epsilon=10^{-6}$.
Runtime rises quickly with~$n$ because the recursive decomposition calls the single-qubit pipeline more often and the synthesized circuits grow with the analytic $T$-count scaling; moving from $10^{-3}$ to $10^{-6}$ increases time only moderately, consistent with the logarithmic dependence of rotation synthesis depth on~$\epsilon$.

\begin{table}[t]
    \centering
    \caption{Unitary synthesis: mean wall time (seconds) per run.}
    \label{tab:unitary-synth-time}
    \begin{tabular}{@{}lcccc@{}}
        \toprule
        $\epsilon$ & $n{=}1$ & $n{=}2$ & $n{=}3$ & $n{=}4$ \\
        \midrule
        $10^{-3}$  & 0.061   & 0.308   & 1.574   & 7.991   \\
        $10^{-6}$  & 0.0873  & 0.47    & 2.283   & 11.322  \\
        \bottomrule
    \end{tabular}
\end{table}

For $n=1$ [panel~(b)], the error decreases as the $T$-count grows in a way consistent with the expected $\black{7\log_2(1/\epsilon)}$ scaling: magnitude approximation contributes $\black{\log_2(1/\epsilon)}$ $T$ gates and the two GridSynth $Z$-rotations each contribute $\black{3\log_2(1/\epsilon)}$ $T$ gates.
For $n=2$ [panel~(c)], the data align with $\black{33\log_2(1/\epsilon)}$ from four single-qubit syntheses, two magnitude approximations, and one GridSynth call in the three-CNOT template.
For $n=3$ and $n=4$ [panels~(d) and~(e)], the error--$T$-count curves steepen with $n$, matching the growth of the analytic $T$-count bound $\black{(21/8 \cdot 4^n - 9/2 \cdot 2^n + 9)\log_2(1/\epsilon)}$.
Overall, the plots support recursive block ZXZ decomposition with magnitude approximation and partial decomposition of intermediate phases to keep the prefactor smaller than full decomposition at every step.

\subsection{Mixed Synthesis Benchmarks}

We benchmark mixed synthesis using the construction and PTM linear program of Section~\ref{sec:mixed} together with the multi-qubit Clifford+$T$ synthesizer (Section~\ref{sec:multi-qubit}).
\black{Perturbations use strength~$\epsilon$ in $\hat{U}=\exp(i\epsilon H_i)$, while the unitary synthesis tolerance is set to~$\epsilon/10$ for numerical stability; \black{for $n$-qubit runs we set $M=2\cdot 4^n$, matching the heuristic $M \gg 4^n$ choice discussed in Sec.~\ref{sec:mixed} (many candidates per target scale).}}
Runs sweep~$\epsilon$ and redraw perturbations independently, and we record the diamond-norm error of the optimized mixture and its total $T$-count.

\begin{figure}[htbp]
    \centering
    \small
    \begin{minipage}[t]{0.48\linewidth}
        \centering
        \includegraphics[width=\linewidth]{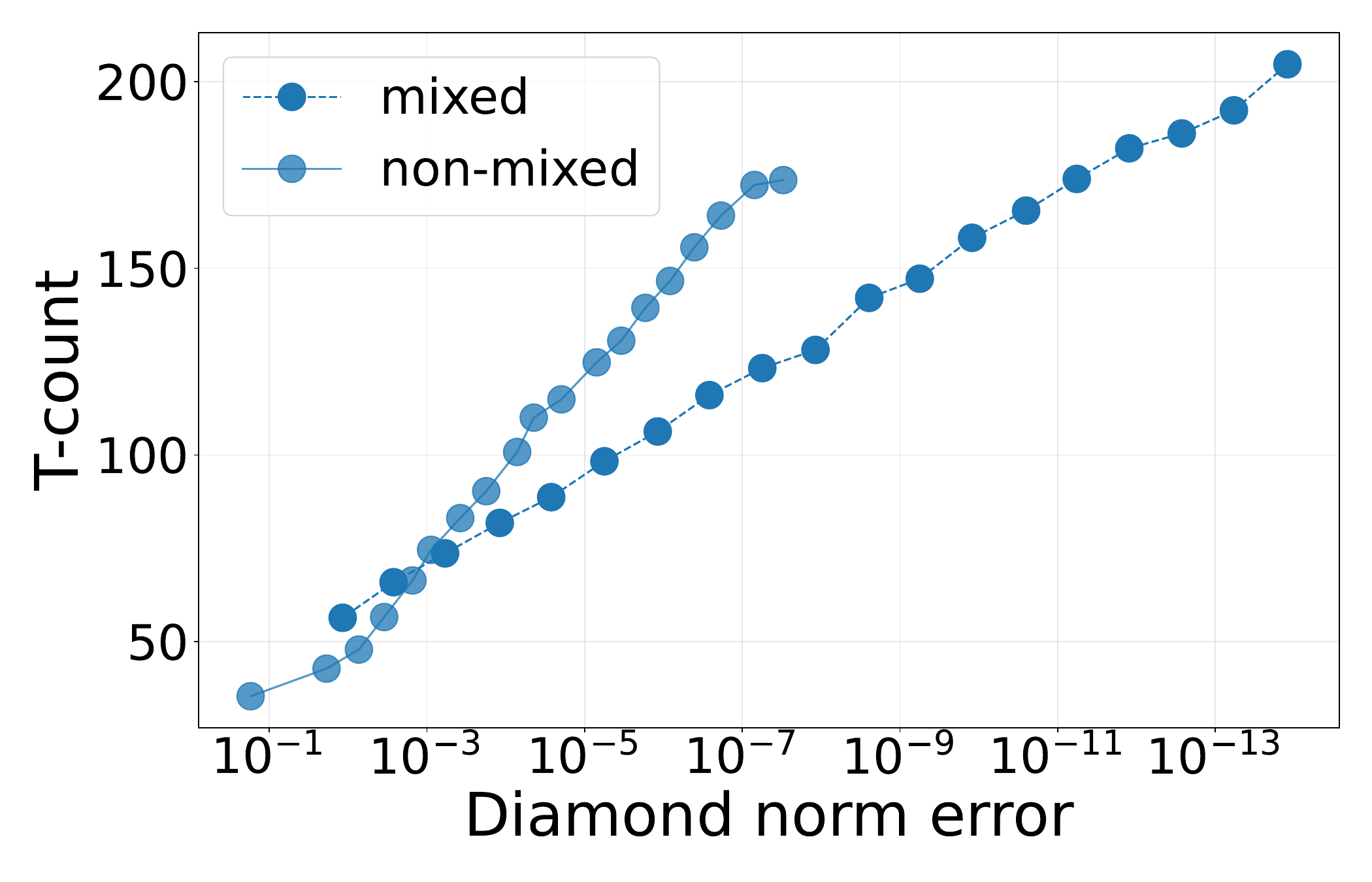}\\[-0.2em]
        {\footnotesize (a) $n=1$}
    \end{minipage}\hfill
    \begin{minipage}[t]{0.48\linewidth}
        \centering
        \includegraphics[width=\linewidth]{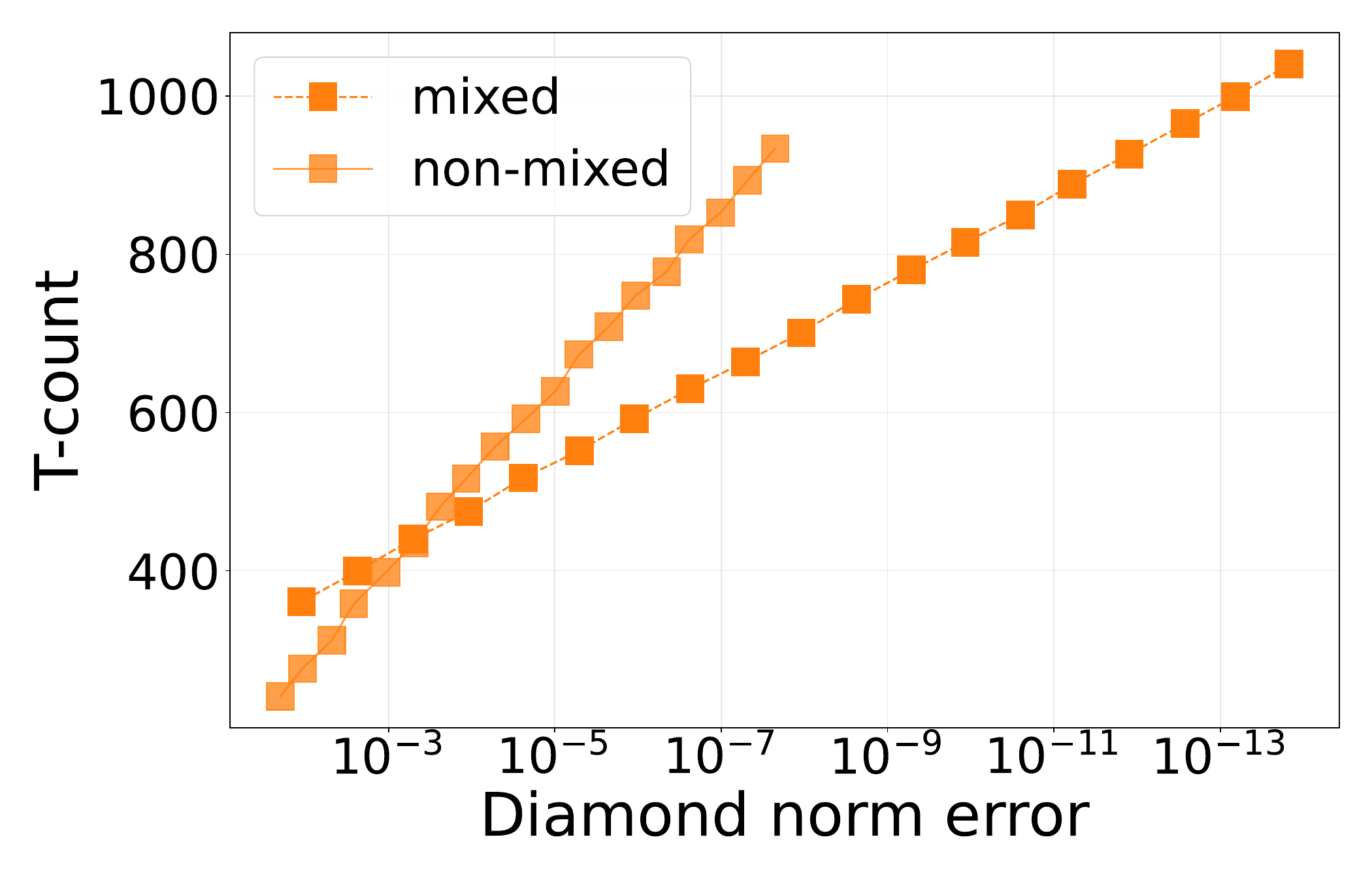}\\[-0.2em]
        {\footnotesize (b) $n=2$}
    \end{minipage}\\[0.6em]
    \begin{minipage}[t]{0.48\linewidth}
        \centering
        \includegraphics[width=\linewidth]{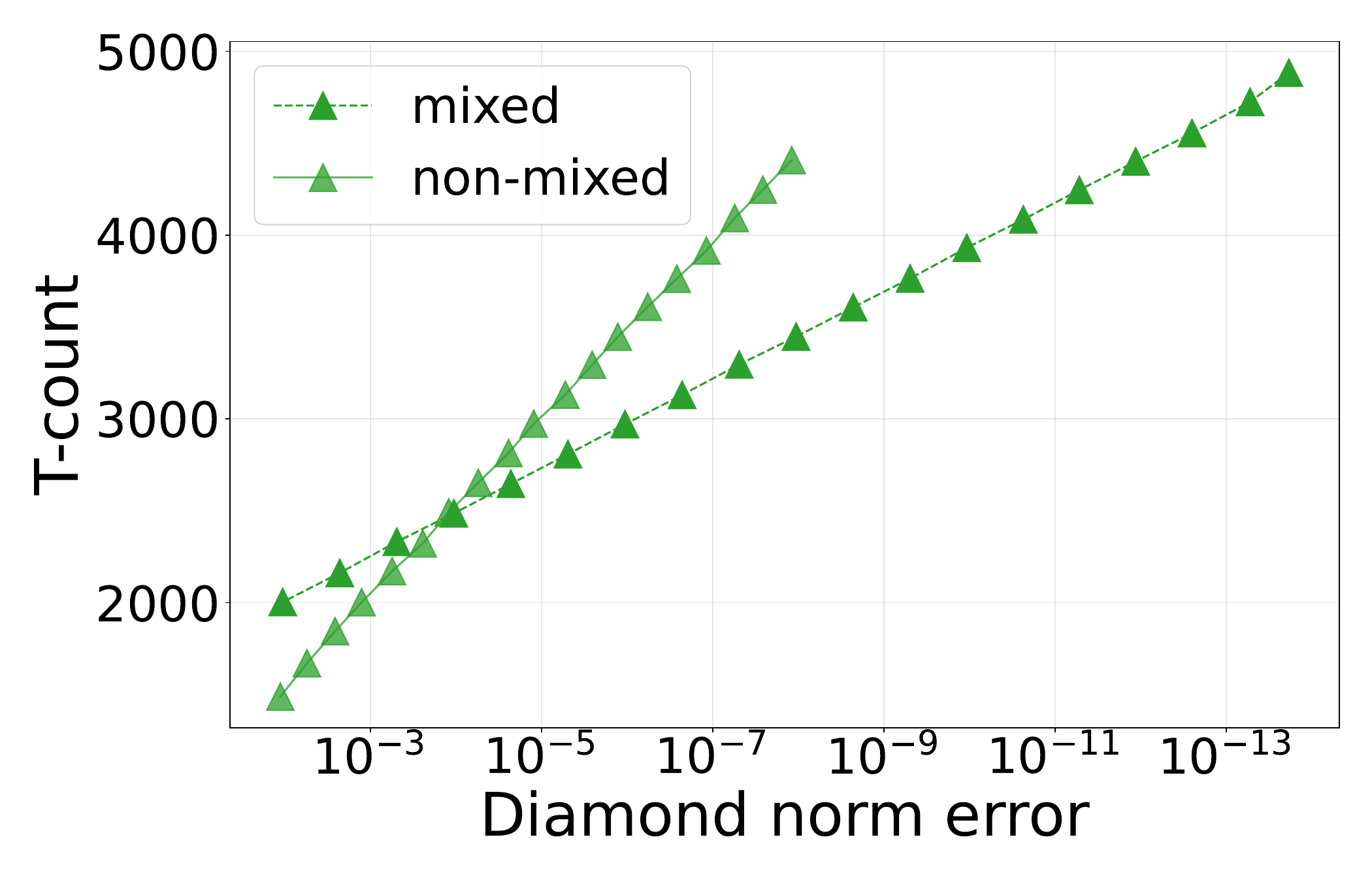}\\[-0.2em]
        {\footnotesize (c) $n=3$}
    \end{minipage}\hfill
    \begin{minipage}[t]{0.48\linewidth}
        \centering
        \includegraphics[width=\linewidth]{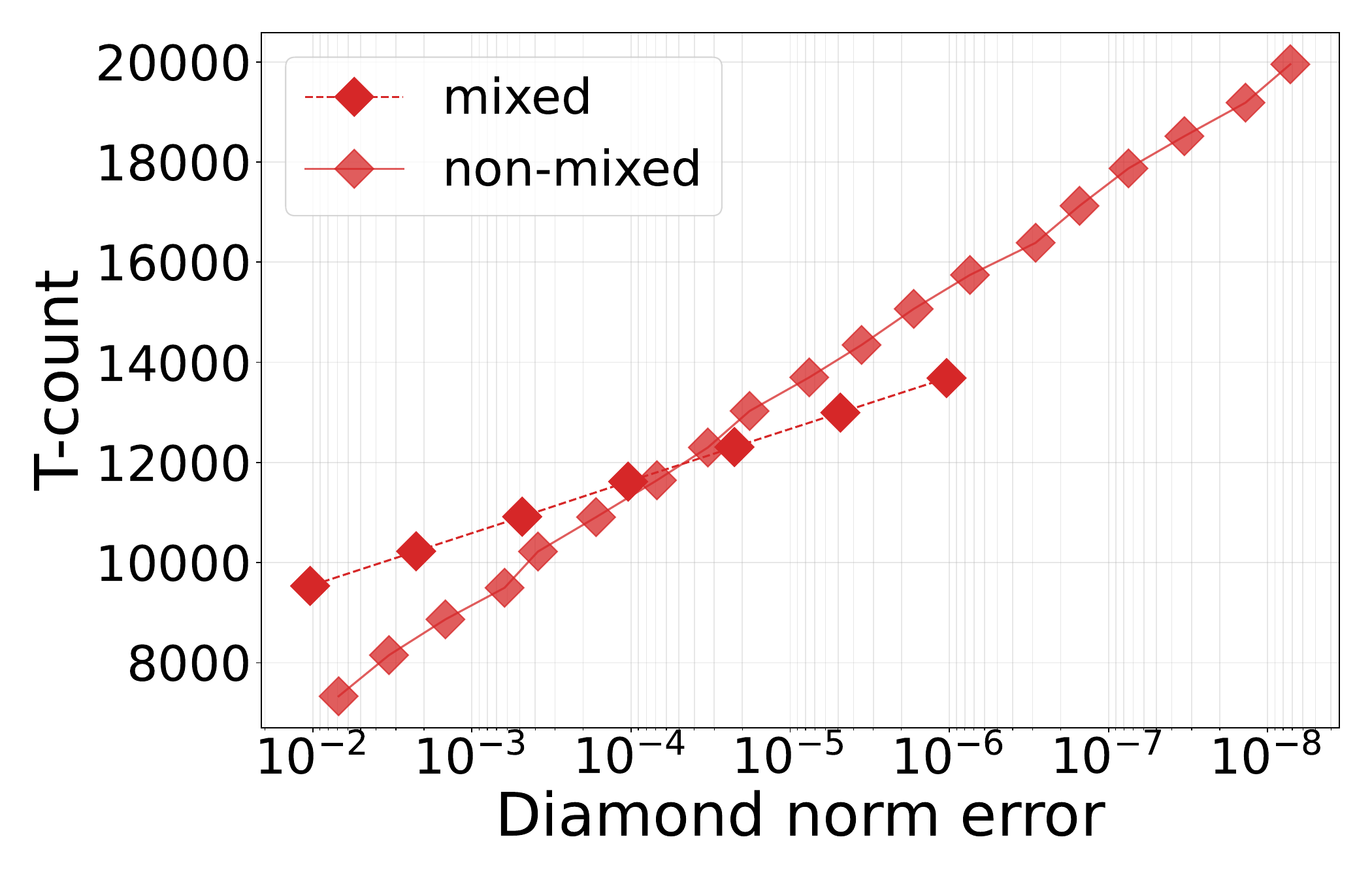}\\[-0.2em]
        {\footnotesize (d) $n=4$}
    \end{minipage}
    \caption{Mixed synthesis benchmarks (error vs.\ $T$-count).
        Each panel fixes the number of qubits $n$; the horizontal axis is the total $T$-count after mixing and the vertical axis is the diamond-norm distance between the target channel and the optimized mixture.
        (a)~$n=1$. (b)~$n=2$. (c)~$n=3$. (d)~$n=4$.
        \black{Panel~(d) omits the smallest-$\epsilon$ regime: at $n=4$ those mixed-synthesis runs exceeded practical wall-clock limits.}}
    \label{fig:mixed-dnorm-tcount}
\end{figure}

\black{Figure~\ref{fig:mixed-dnorm-tcount} summarizes these tradeoffs.
    Across the panels ($n=1$--$4$), we find that mixed synthesis in the small error limit halves the $T$-count of unitary synthesis at comparable diamond-norm error, consistent with Lemma~\ref{lem:mixed-synth}: if the candidates form an $\epsilon$-net in diamond \black{norm error}, the optimized mixture can achieve $O(\epsilon^2)$ residual error, so the leading $\log_2(1/\epsilon)$ cost per decade of target error is halved.
    Mixed synthesis also incurs a larger additive $T$-cost: the candidates must be synthesized with accuracy well below the perturbation scale~$\epsilon$ used to generate them, which adds $T$ gates whose contribution is only weakly reduced when~$\epsilon$ is large.
    At small~$\epsilon$ the halved leading term dominates and mixed synthesis clearly outperforms unitary synthesis; at large~$\epsilon$ this additive cost can dominate, so mixed runs need not use fewer total $T$ gates for comparable accuracy.}

\begin{figure}[htbp]
    \centering
    \small
    \begin{minipage}[t]{0.48\linewidth}
        \centering
        \includegraphics[width=\linewidth]{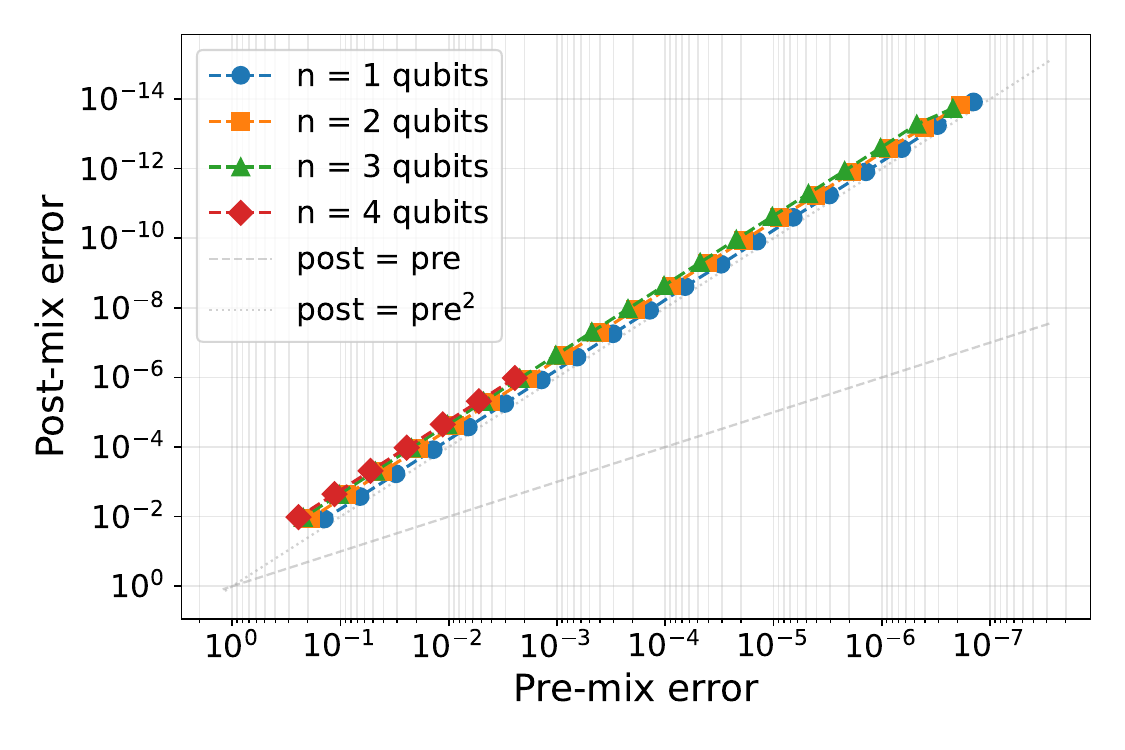}\\[-0.2em]
        {\footnotesize (a) Horizontal: pre-mix; vertical: post-mix.}
    \end{minipage}\hfill
    \begin{minipage}[t]{0.48\linewidth}
        \centering
        \includegraphics[width=\linewidth]{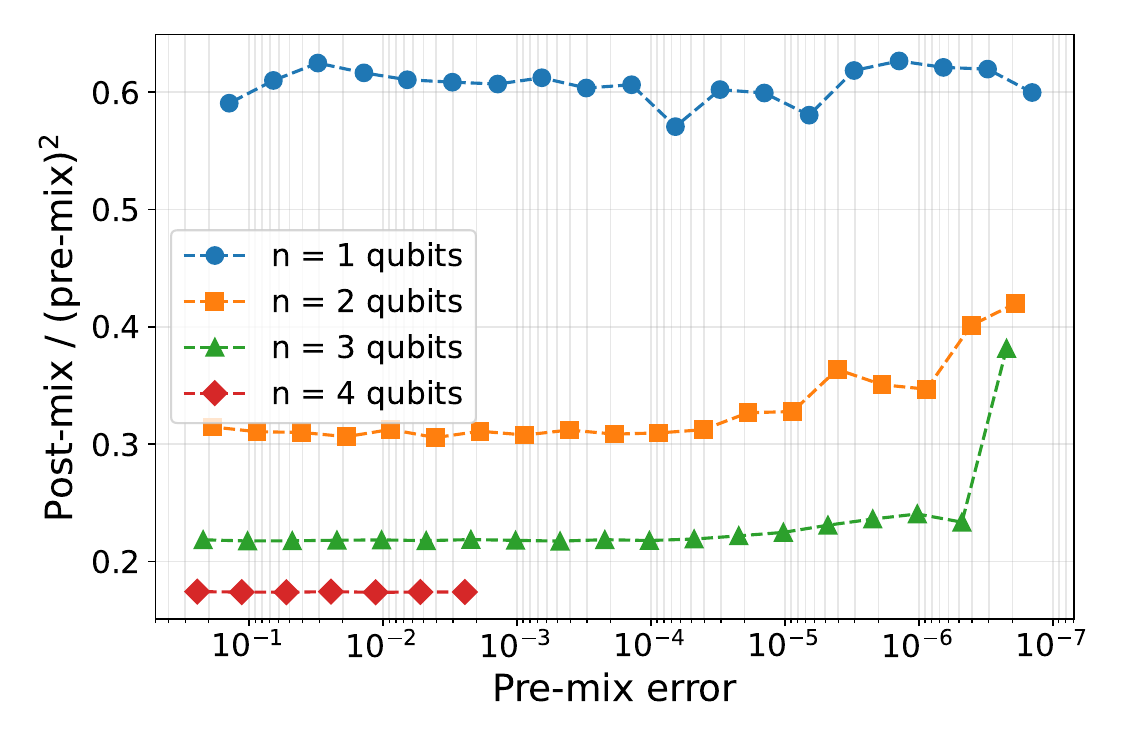}\\[-0.2em]
        {\footnotesize (b) Horizontal: pre-mix; vertical: $\text{post-mix}/(\text{pre-mix})^2$.}
    \end{minipage}
    \caption{Mixed synthesis (Section~\ref{sec:mixed}): each marker is one run at some~$\epsilon$ and~$n$.
        Pre-mix is the diamond-norm distance from the target to a typical synthesized perturbed unitary before LP mixing; post-mix is the diamond-norm distance from the target to the optimized mixture channel after mixing.
        (a)~Horizontal axis: pre-mix; vertical axis: post-mix.
        (b)~Horizontal axis: pre-mix; vertical axis: post-mix divided by $(\text{pre-mix})^2$.}
    \label{fig:mixed-pre-post}
\end{figure}

Figure~\ref{fig:mixed-pre-post}(a) places post-mix well below pre-mix for comparable horizontal values, and on log--log axes the trend is consistent with post-mix error scaling approximately as $(\text{pre-mix})^2$, as predicted by Lemma~\ref{lem:mixed-synth} when the candidates are close to the target in diamond norm.
Panel~(b) removes this quadratic factor by dividing the vertical coordinate by $(\text{pre-mix})^2$; the separation between curves with different~$n$ 
\black{can be empirically understood as}
mixture error picking up an additional factor of order $1/(2n)$ relative to that $\epsilon^2$-type baseline.
\black{We leave the mathematical proof for such an improvement as a future work.}

\section{Conclusion}\label{sec:conclusion}



We have presented \texttt{pygridsynth}, an open-source Python library for ancilla-free approximate synthesis over the Clifford+$T$ gate set. The package combines arbitrary-precision implementations of established single-qubit primitives with a recursive workflow for multi-qubit unitary approximation. This design makes it possible to study high-precision Clifford+$T$ compilation within a single Python-native environment that is suitable both for practical integration with quantum software stacks and for systematic resource benchmarking in the fault-tolerant regime.


Various future extensions and developments can be envisioned.
A first natural extension is to move beyond the ancilla-free setting and incorporate adaptive synthesis with ancilla qubits, mid-circuit measurement, and classical feed-forward. Recent work on multi-qubit Toffoli gates shows that once randomized Clifford+$T$ implementations are allowed, the required $T$-count can drop to $O(\log(1/\epsilon))$, whereas any purely unitary constant-error implementation still requires $\Omega(n)$ $T$ gates \cite{gosset2025multi}. This suggests that non-unitary resources can qualitatively change the multi-qubit synthesis landscape. Extending \texttt{pygridsynth} to support ancilla-assisted and adaptive backends would therefore be valuable both practically and conceptually: practically, because such protocols may reduce fault-tolerant resource costs for important controlled operations; conceptually, because they would clarify where the true boundary lies between ancilla-free unitary synthesis, mixed synthesis, and fully adaptive compilation.

A second priority is to expose optional backends that trade classical runtime for provable $T$-optimality. Recent work by Morisaki, Sano, and Akibue gives ancilla-free single-qubit Clifford+$T$ synthesis algorithms that are optimal in $T$-count, with asymptotic $T$-count of $3\log_2(1/\epsilon)+o(\log(1/\epsilon))$ gates for most SU(2) gates, albeit with classical runtimes that are exponential in the input size \cite{morisaki2025optimal}. Integrating such methods as an optional backend would let users choose between fast practical compilation and slower but provably optimal synthesis on selected high-value instances. More broadly, it remains open whether multi-qubit mixed synthesis can be pushed all the way to the relevant lower bounds, rather than merely exhibit quadratic suppression empirically. The optimal-convex-approximation viewpoint \cite{akibue2024probabilistic} and recent work on error crafting in mixed gate synthesis \cite{Yoshioka2025error} suggest that there is still room to exploit structure in the remnant error. 
Understanding whether stronger optimization procedures, more structured coverings of the target set, or adaptive ancilla-assisted protocols can saturate these bounds is, in our view, one of the most interesting directions opened by the present work.


\section*{Acknowledgements}
The authors would like to thank Seiseki Akibue, Shelly Garion, Ali Javadi-Abhari for fruitful discussions.
N.Y. is supported by JST Grant Number JPMJPF2221, JST CREST Grant Number JPMJCR23I4, IBM Quantum, Google AI Quantum, JST ASPIRE Grant Number JPMJAP2316, JST ERATO Grant Number JPMJER2302,  JST [Moonshot R\&D] [Grant Number JPMJMS256J].

\bibliographystyle{unsrtnat}
\bibliography{reference}

\appendix
\section{Perturbation-Based Unitaries}
\phantomsection
\label{sec:perturbation-appendix}

The function \texttt{process\_unitary\_approximation\_parallel} (or its sequential variant) generates a set of perturbed unitaries:
\begin{equation}
    U_i = \hat{U} \exp(i \epsilon H_i),
\end{equation}
where $H_i$ are randomly generated Hermitian operators and $\epsilon$ is the error tolerance parameter.
These perturbed unitaries are then approximated using the multi-qubit synthesis algorithm described in Section~\ref{sec:multi-qubit} (\texttt{approximate\_multi\_qubit\_unitary}), resulting in efficiently synthesizable Clifford+$T$ circuits.
Note that the approximation error introduced by the multi-qubit synthesis algorithm is also included in the perturbation, so the actual error between the target unitary $\hat{U}$ and each synthesized unitary $U_i$ may be slightly larger than $\epsilon$.

The Hermitian operators $H_i$ are generated using the function \texttt{get\allowbreak\_random\allowbreak\_hermitian\allowbreak\_operator} as follows:
For an $n$-qubit system with Hilbert space dimension $d = 2^n$, the algorithm samples $M$ points uniformly from the unit sphere $S^{d^2-2}$ in $(d^2-1)$-dimensional space.
These points are then relaxed using a repulsive force algorithm to ensure better distribution: for each point $x_i$, a repulsive force $F_i = \sum_{j \neq i} (x_i - x_j) / (\norm{x_i - x_j}^2 + \delta)$ is computed, projected onto the tangent space of the sphere, and the point is updated iteratively.
Each relaxed point is then converted to a Hermitian matrix $H$ of dimension $d \times d$ by interpreting it as a Pauli vector representation and converting it to a density matrix using the inverse Pauli-to-state mapping.
This process ensures that the set of Hermitian operators $\{H_i\}$ provides diverse perturbations, forming an $\epsilon$-net around the target unitary $U$ that enables the subsequent optimization to achieve quadratic error suppression.

\end{document}